\documentclass[12pt]{iopart}
\usepackage{iopams}
\usepackage{url}
\usepackage{graphicx,color}
\usepackage{verbatim}

\usepackage{color}
\usepackage[usenames,dvipsnames,svgnames,table]{xcolor}
\usepackage{hyperref}
\hypersetup{
	colorlinks   = true,
	citecolor    = blue
}

\expandafter\let\csname equation*\endcsname\relax
\expandafter\let\csname endequation*\endcsname\relax
\usepackage{amsmath}
\usepackage{epstopdf}

\usepackage{cleveref}

\crefname{equation}{}{}
\crefname{figure}{Figure}{Figures}
\crefname{appendix}{}{}

\graphicspath{{./fig/}}

\newcommand{\QuenchedAverage}[1]{\overline{#1}}
\newcommand{\PartitionFunc}[1]{\overline{Z_{#1}^P }}
\newcommand{\Action}[1]{\mathcal{F}_{#1}}
\newcommand{\ActionCoeff}[2]{\mathcal{F}_{#1}^{(#2)} }
\newcommand{\SquareBracket}[1]{\left[ #1 \right]}
\newcommand{\Parenthesis}[1]{\left( #1 \right)}
\newcommand{\Avr}[1]{\left \langle #1  \right \rangle}

\newcommand{\SumConstraintOneSite}[1]{\{\sigma(t)\}^{#1}}

\newcommand{\Measure}[1]{\Tr_{\lambda, \sigma}^{#1} \,}
\newcommand{\MeasureSite}[1]{\tr _{\lambda, \sigma}^{#1}\,}

\let\a=\alpha   
\let\e=\varepsilon  \let\h=\eta \let\k=\kappa
\let\l=\lambda    
\let\s=\sigma   
   
   \let\P=\Pi
\let\Si=\Sigma   
\let\ee=\epsilon  \let\th=\theta \let\io=\infty

\def\DD{{\cal D}}

\def\ul{\underline}

\def\bs{\boldsymbol{\sigma}}

\begin{document}
\title{On the number of limit cycles in asymmetric neural networks}

\author{Sungmin Hwang$^1$, Viola Folli$^2$, Enrico Lanza$^3$, Giorgio Parisi$^4$, Giancarlo Ruocco$^{2,4}$, Francesco Zamponi$^5$}

\address{$^1$ LPTMS, Universit\'e Paris-Sud 11, UMR 8626 CNRS, B\^at. 100, 91405 Orsay Cedex, France\\
		$^2$ Fondazione Istituto Italiano di Tecnologia (IIT), Center for Life Nano Science, Viale Regina Elena 291, I00161 Roma, Italy \\
		$^3$ Dipartimento di Biotecnologie Cellulari ed Ematologia, Sapienza Universit\`a di Roma, P.le A. Moro 2, 00185 Roma, Italy \\
		$^4$ Dipartimento di Fisica, Sapienza Universit\`a di Roma, P.le A. Moro 2, 00185 Roma, Italy \\
		$^5$ Laboratoire de physique th\'eorique, D\'epartement de physique de l'ENS, \'Ecole normale sup\'erieure, PSL  Research University, Sorbonne Universit\'es, CNRS, 75005 Paris, France 		}

\vspace{10pt}
\begin{indented}
\item[]\today
\end{indented}
\begin{abstract}
The comprehension of the mechanisms at the basis of the functioning of complexly interconnected networks represents one of the main goals of neuroscience. 
In this work, we investigate how the structure of recurrent
connectivity influences the ability of a network to have
storable patterns and in particular limit cycles, by modeling a recurrent neural
network with McCulloch-Pitts neurons as a
content-addressable memory system. 

A key role in such models is played by the connectivity matrix, which,
for neural networks, corresponds to a schematic representation of the ``connectome":
the set of chemical synapses and electrical junctions among neurons.
The shape of the recurrent connectivity matrix plays a
crucial role in the process of storing memories.
This relation has already been exposed by the work of Tanaka and Edwards, which presents a theoretical approach to evaluate the mean number of fixed points in a fully connected model at thermodynamic limit. 
Interestingly, further studies on the same kind of model but with a finite number of nodes have shown how the symmetry parameter influences the types of attractors featured in the system.
Our study extends the work of Tanaka and Edwards by providing a theoretical evaluation of the mean number of attractors of any given length $L$ for different degrees of symmetry in the connectivity matrices.
\end{abstract}
\vspace{2pc}

\submitto{\JSTAT}

\section{Introduction}

Understanding the collective functioning, the emerging properties and cognitive processes of a large network of 
complexly interconnected neurons on the basis of local activity and neuronal circuitry represents one of the primary goals of neuroscience \cite{Sporns2013,Park2013,Martin2001,Pascual2004,Yamashita2015,Shultz2016}.
The mammalian brain contains billions of neurons and hundred trillions of synapses and the complexity of the biological neural networks increases exponentially with dimension, being higher brain systems working on apparently quasi-segregated areas that indeed are complexly connected and integrated with each other \cite{Sporns2013}.
The comprehension of the way how a brain works is  one of the most fascinating problems in modern science, to the extent that some authors claim that \textit{``...the capacity of any explaining agent must be limited to objects with a structure possessing a degree of complexity lower than its own.
If this is correct, it means that no explaining agent can ever explain objects of its own kind, 
or of its own degree of complexity, and, therefore, that the human brain can never fully explain its own operations."} \cite{Hayek1963}.
Being this {\it prima facie} hypothesis true or not, the problem presents such a high level of complexity that the use of (over)simplified models is unavoidable.
Indeed, controlling the global behavior of an artificial neural network and the resulting collective adaptive behavior and information processing at the level of local structural connectivity and synaptic asymmetry may shed light on the functioning of living nervous systems.

In this work, we investigate how the structure of recurrent
connectivity influences the ability of the network to have
storable patterns, and in particular limit cycles of a given length $L$.
To this aim, we model a recurrent neural
network with McCulloch-Pitts neurons~\cite{McCulloch1943}  as a
content-addressable memory system~\cite{Hopfield1982}.
In a recurrent neural network, the information is stored
nonlocally and the memory retrieval process is associated with
complex neuronal activation patterns (attractors) encoding
memory events.
The strength of these patterns fixes the 
ability to quickly recall memories of a specific event.
The architecture of the connectivity matrix itself determines the
clustering operation of the set of data inputs, reducing the complexity of 
the $ N $-dimensional initial problem (many-to-few mapping).
The weights of the connections between cells are self-organized on the
basis of the set of input patterns, and the asymptotic solution
of network dynamics represents the response
of the network to a given stimulus with which the initial
condition is identified.

The Hopfield model \cite{Hopfield1982,Amit1985},
and the idea that information is stored via attractor states
has proven to be a powerful conceptual tool in neuroscience.
Indeed, there is some experimental support for discrete attractors
in the patterns of activity of hippocampal cells during spontaneous
activity in rodents \cite{Pfeiffer2015} or persistent activity in
monkeys during tasks \cite{Fuster1971,Miyashita1988}.
Furthermore, the Hopfield model may benefit from the analogy between neural networks
and spin systems (for which some interesting results have already been obtained) since both models refer to a network of elementary units, whose dynamics depend on the interaction of neighboring elements.
A key role in these models is played by the connectivity matrix $J_{ij}$.
For neural networks, the matrix $J_{ij}$ is a schematic representation of the ``connectome":
the set of chemical synapses and electrical junctions among neurons.
The shape of the recurrent connectivity determined by $J_{ij}$ plays a
key role in the process of storing memories. 

Such dependency has been explored in the work of Tanaka and Edwards \cite{Tanaka1980},
which presents a theoretical approach to evaluate the mean number of fixed points in a 
random ensemble of
fully connected Ising spin glass models at thermodynamic limit. 
A network of $N$ binary neurons, encoded as spin (binary) variables
${\sigma_i \in \{-1,1 \}}$ 
($i=1,\dots,N$),
presents $2^N$ different possible states or ``firing 
patterns'' $\bs(t)=\{\sigma_1(t),\sigma_2(t),\dots,\sigma_N(t)\}$.
At each time step, the evolution rule updates synchronously all nodes according to the following rule:
\begin{equation}
\label{eq1}
\sigma_i(t+1)=\textrm{sgn}\Big(\sum\limits_{j=1}^N J_{ij}\sigma_j(t)\Big),
\end{equation}
where $\textrm{sgn}(x)$ is the sign function.
The matrix element $J_{ij}$  represents the strength of the connection
between node $i$ and $j$ and is assumed to be a quenched random variable drawn from a fixed distribution with zero mean. 
Because the chosen dynamics is deterministic, each state is univocally connected
to another one: this results in a deterministic path in the state space towards the
corresponding attractor.
Indeed, as the state-space is finite, the dynamics necessarily reaches a ``final" state,
that can be either a fixed point or a limit cycle of a certain length $L$ (1$\le L \le 2^N$).

Interestingly, further studies on the same kind of model but with a finite number of nodes have shown numerically how the symmetry parameter influences the types of attractors featured in the system \cite{Gutfreund1988}.
One way of quantifying the symmetry of the connection's strengths $J_{ij}$ is through the symmetry parameter $\eta$, which corresponds to the following value: 
\begin{align*}
\eta=\frac{\langle J_{ij} J_{ji} \rangle}{\langle J_{ij}^2 \rangle}. 
\end{align*}
With this definition, $\eta=1$ represents symmetric connectivity matrices, $\eta=0$ asymmetric ones, while $\eta=-1$ refers to antisymmetric matrices.
The main finding in this case is linked to a transition at $\eta_c\approx 0.5$~\cite{Bastolla1998}.
For $\eta>\eta_c$, these systems feature mainly  fixed points or limit cycles of length 2 whose number increases exponentially with $N$, while for systems with $\eta<\eta_c$ the typical length of limit cycles increases exponentially with $N$ and the dynamics is chaotic. 
Note that the transient time $\tau$~\cite{Gutfreund1988,Nutzel1991},
which is the time needed by the system to reach the corresponding limit cycle from a randomly chosen initial state,
is exponential ($ \tau \sim e^{Ng(\eta)}$) for any $\eta<1$, and it only becomes polynomial at $\eta=1$~\cite{Bastolla1998}.
An additional dynamical transition may be seen in the chaotic regime around $\eta_d\approx 0.33$.
For $\eta_d<\eta<\eta_c$, the number of limit cycles of length 2 is exponentially high, but with vanishing basins of attraction.
For $0<\eta<\eta_d$, exponentially long limit cycles have dominating basins of attraction \cite{Bastolla1998}.
Other works investigated the existence of transitions in generalizations of this system, e.g. adding noise \cite{Molgedey1992,Schuecker2016} and dilution \cite{Tirozzi1991}, with the possible use of other characteristic values that may be linked to transitions, like the gain function \cite{Sompolinsky1988,Crisanti1990} or self-interaction \cite{Stern2014}. Interestingly, recent numerical studies demonstrated how the dilution of a fully asymmetric  network leads to an increase in the complexity \cite{Folli2017}. The analytical study of the dynamics of these networks was initiated in~\cite{GardnerE.1987,Derrida1986,DerridaB.1987}.

For practical purposes, in order to numerically construct a $J$ matrix with a given symmetry, we introduce a different symmetry parameter.
Specifically we exploit the following representation of the matrix elements: 
\begin{equation}
\label{eq2}
J_{ij}=\left(1-\frac{\epsilon}{2}\right)S_{ij}+\frac{\epsilon}{2}A_{ij},
\end{equation}
where $S_{ij}$ and $A_{ij}$ are
symmetric and antisymmetric random matrix elements respectively (with $S_{ji}$=$S_{ij}$, $A_{ji}$=$-A_{ij}$, while $S_{ii}$=$A_{ii}$=0),
independently extracted from a fixed distribution $P(x)$. As will be explicitly shown later, our main results will be largely 
independent of the choice of $P(x)$ as long as 
it does not depend on the size of system $ N $. For the numerical simulations, however, we tested our results for a 
Gaussian, a uniform and a binary distribution of $J_{ij}$.
The aforementioned symmetry parameter $\eta$ widely used in the literature is related to $\epsilon$
by
\begin{equation}
\label{eq3}
\eta=\frac{1-\epsilon}{1-\epsilon+\frac{\epsilon^2}{2}}.
\end{equation}
For $\epsilon = 0$ (or equivalently $\eta=1$) all nodes interact symmetrically with each other, whereas for $\epsilon=1$ ($\eta=0$) and $\epsilon=2$ ($\eta=-1$) their interaction is asymmetric and antisymmetric respectively. 

In this paper, we focus on the problem of counting the number of limit cycles of length $L$, irrespectively of their basins of attraction.
We extend the work of Tanaka and Edwards \cite{Tanaka1980}, 
in which the exponential growth rate $\Sigma_1$ of the number of fixed points was computed. 
Specifically, we develop a framework based on the one of Ref.~\cite{GardnerE.1987} 
that allows us to determine, for different degrees of symmetry $\eta$ in the connectivity matrices, 
the average number of attractors of any length $L$ of the form $n_L  \sim (A_L/L) e^{N \Sigma_L}$. 
These results are then used to support the existence of a transition of the type discussed above. 
Besides, our approach provides additional information on the cycle structures such as the
overlap parameters between configurations forming a cycle. 
Finally, thanks to the fact that our formalism is exact, this can also be used to verify the approximations and assumptions made in the analytical arguments employed in \cite{Bastolla1998}.

Our manuscript is organized as follows.
In \cref{sec:method}, we first present the results of numerical simulations to provide the overall picture of the dynamics.
Next, we develop a statistical mechanics formalism that computes the number of attractors of a given length $L$. 
This translates our problem into an optimization problem over a finite set of variables.
In \cref{sec:result},
 we analytically determine the exponential growth rates $\Sigma_L$ for $L=1, L=2$ and $L=3$ by numerically solving the corresponding optimization problems.
Then, we move on to the case of arbitrary longer cycle lengths $L$ in the vicinity of $\eta =0$ where a perturbative approach is valid. 
Finally in \cref{sec:discussion}, we discuss the implications of our results, especially in terms of the transition to chaos.


\section{Methods}
\label{sec:method}

We consider a network of $N$ binary neurons ${\sigma_i \in \{-1,1 \}}$ 
evolving according to Eq.~\eqref{eq1}, with quenched couplings $J_{ij}$ constructed as in Eq.~\eqref{eq2} using a distribution $P(x)$ to be specified in the following,
and a symmetry parameter $\ee$.
It is worth to note that, because the connectivity matrix appears in the dynamical equation only as the argument of the $\mathrm {sign}$ function, any scaling $\mathbf J \rightarrow \alpha \mathbf J$ (with $\alpha > 0$) does not alter the dynamics.
Note that we set $J_{ii}$=0
(no autapses) to exclude chaotic behavior where the system is
characterized by an extreme sensitivity to initial conditions and
two nearly identical starting points will reach different
attractors \cite{Sompolinsky1988}.

In this paper, we mainly focus on the long-time properties of the dynamics, i.e., the statistical 
properties of periodic points (or limit cycles) of the dynamics.
The non-existence of Hamiltonian implies 
that cycles with any length can exist in the system.
As an exception, it can be shown that there exists an energy function at  $\epsilon=0$ only allowing cycles of length $L=1$ or $L=2$ \cite{Bastolla1998}. Similarly, at $\epsilon=2$ only cycles of length $L=4$ exist.
In the following, we present a general formalism, based on a slight modification of the one developed in~\cite{GardnerE.1987}, 
that computes the average of the number of $L$-cycles in the limit $N\rightarrow\infty$.

\subsection{Numerical simulations}

To evaluate numerically the average properties of the limit cycles of the networks, we randomly generate a 
statistically significant number of realizations of connectivity matrices $ \mathbf J $'s with equal symmetry properties of the same size.
Once the connectivity matrix has been generated for a given pair $(N,\epsilon)$, 
we evolve all $2^N$ initial conditions. Because the configuration space is finite and the dynamics are deterministic,
after a transient time, the system evolves towards a fixed point or a limit cycle.
The algorithm works through a many-to-few mapping connecting $2^N$ initial patterns to the corresponding attractors for each realization of the connectivity matrix. 
The evolution paths are distributed on a number ($n (\mathbf J)$) of oriented graphs each one containing one attractor \cite{Diestel2005}.


For each graph $k$ ($k=1\dots n (\mathbf J))$, we measure the length $L_k$ of its attractor (fixed point, $L=1$, or limit cycles, $L \ge 2$).
Thus, we are readily able to evaluate the number of $L$-cycles, i.e., $n_L (\mathbf J) = \sum_{k=1}^{n(\mathbf J)} \delta_{L, L_k}$. 
After processing a statistically significant number of realizations, we may compute 
 the average number of cycles $\bar n = \overline {n(\mathbf J)} $ and the average cycle length, $\bar L$, as $\bar L = \overline {\sum_L L n_L(\mathbf J) /  n(\mathbf J) } $.
Here, the overline $\QuenchedAverage{(\cdots)}$ is used to denote the average over realizations of $\mathbf{J}$. 
The explored region of $N$ ranges from 8 to 20 for Gaussian couplings and $N$ from 8 to 32 for binary couplings, while the sampling of $\epsilon$ covers the [0,1] range with 0.05 spacing and 0.01 spacing from the critical region (where $\epsilon\approx 0.7$) up to 1.
Other quantities of interest include the size of basins of attraction and the average distance
between a generic state and the corresponding attractor, which are not discussed in the present paper.

\begin{figure}[t]
	\begin{center}
	\includegraphics[width=.75\textwidth]{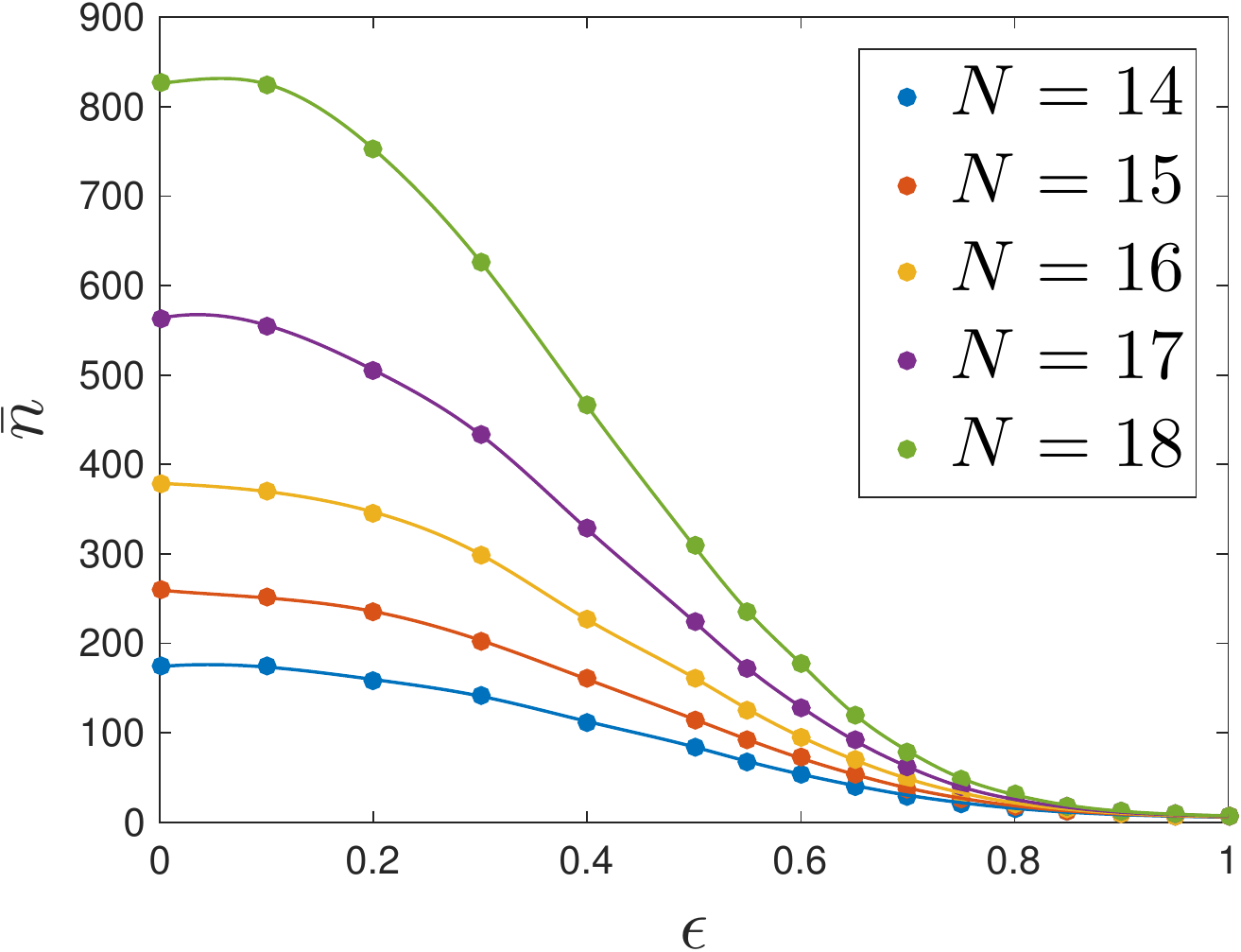}
	\caption{ Mean number of the total amount of limit cycles per matrix, $\bar n$,  as a function of $\epsilon$ for systems of different sizes ($N=14,15,16,17$ and $ 18 $).  Overall, the number of independently generated random matrices for each $\epsilon$ value ranges from 100,000 at $N$=14 to 1,000 at $N$=18. }
	\label{simulazioni1}
	\end{center}
\end{figure}

\begin{figure}[t]
	\begin{center}
	\includegraphics[width=.75\textwidth]{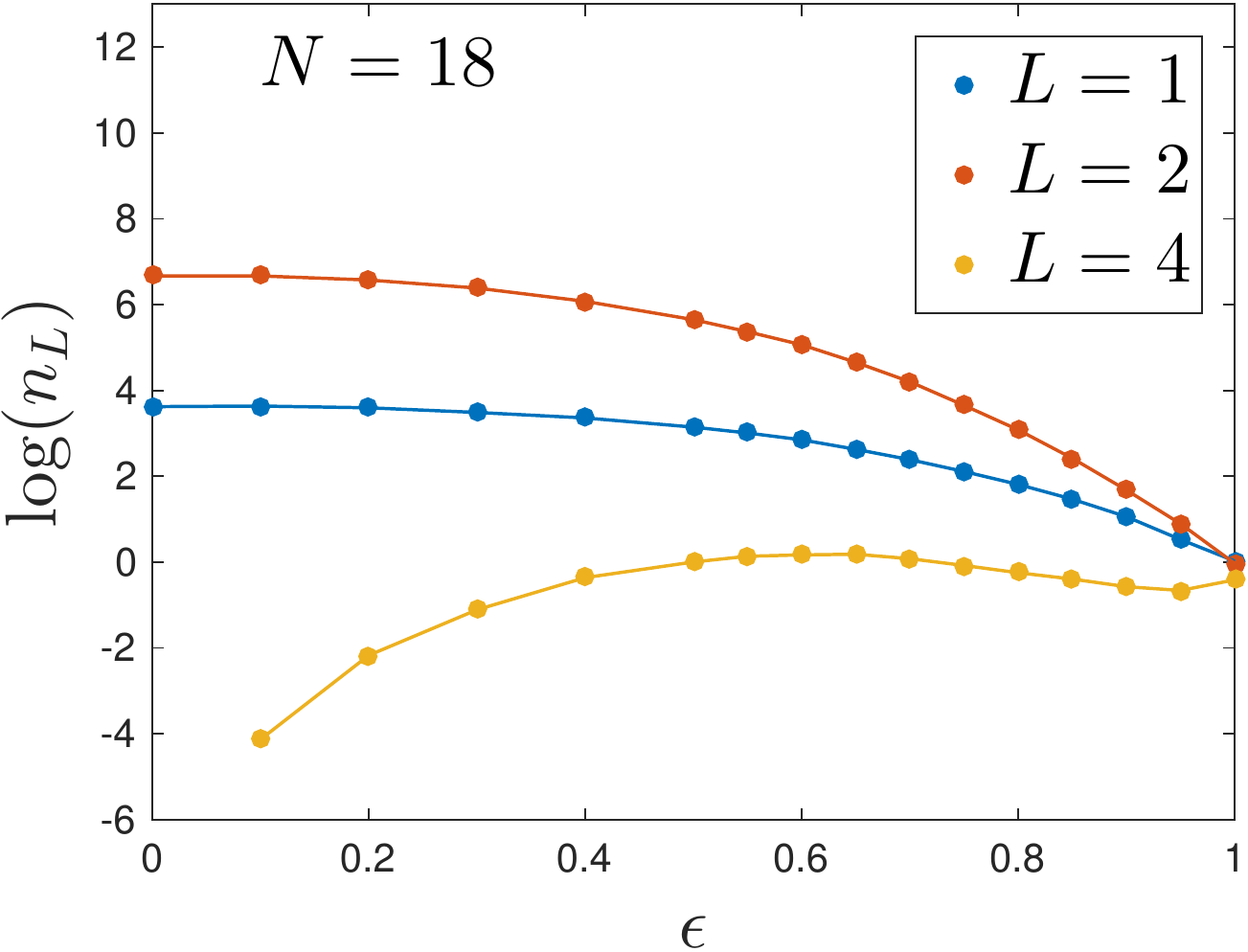}
	\caption{ Mean number of limit cycles of various lengths, $n_L = \overline{n_L(\mathbf{J})}$,  ($L=1,2$ and $ 4 $) in logarithmic scale as a function of $\epsilon$ for systems with 18 nodes. \label{simulazioni2}
	}
	\end{center}
\end{figure}

\begin{figure}[t]
	\begin{center}
	\includegraphics[width=.75\textwidth]{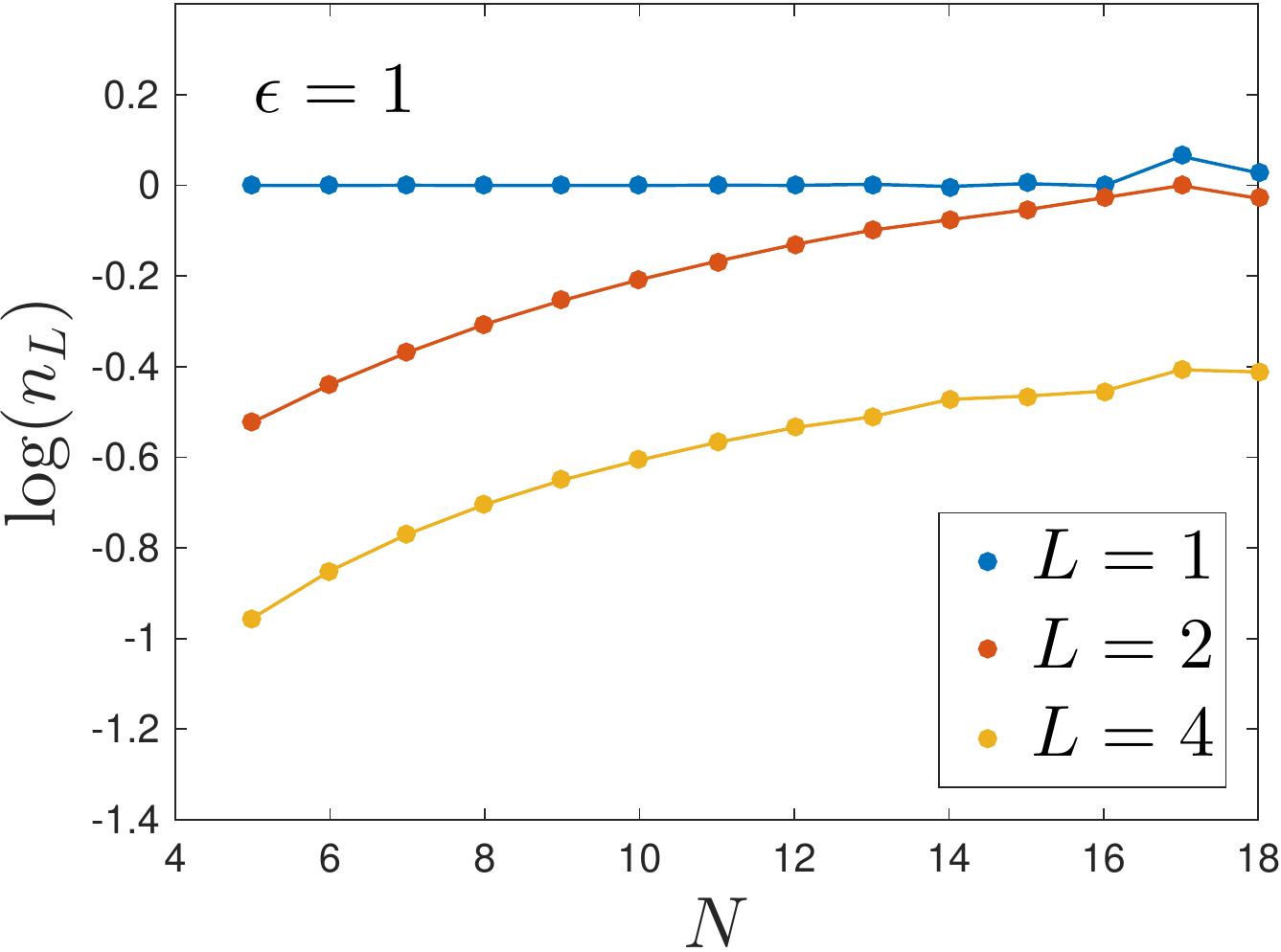}
	\caption{ Mean number of limit cycles of various lengths, $n_L = \overline{n_L(\mathbf{J})}$,  ($L=1,2$ and $ 4 $) in logarithmic scale as a function of $N$ for systems with symmetry parameter $\epsilon=1$.   \label{simulazioni3}
	}
	\end{center}
\end{figure}

Figures \ref{simulazioni1}, \ref{simulazioni2}, and \ref{simulazioni3} show examples of results obtained through our simulations for Gaussian couplings: \cref{simulazioni1} reports the results obtained for the mean total number of limit cycles as a function of $\epsilon$ for various system sizes $ N $, while \cref{simulazioni2} shows the mean number of limit cycles of length 1, 2 and 4 as a function of $\epsilon$ for systems with $ N=18 $. Finally,  \cref{simulazioni3} shows the average number of limit cycles of length 1, 2 and 4 as a function of $ N $, for systems with symmetry parameter $\epsilon=1$.

\subsection{Basic theoretical formalism}

Our formalism follows closely the one of~\cite{GardnerE.1987}, with some adaptation to the problem of interest here.
Given two spin configurations $\bs, \bs'$, let us first ask whether $\bs'$ is the one step evolution of $\bs$ according to \cref{eq1}.
To answer this question, we define a corresponding indicator random variable $w(\bs, \bs')$ which is one if such event occurs and zero otherwise.
A convenient representation of this indicator variable is obtained by observing that
$\bs'$ is the evolution of $\bs$ under \cref{eq1} if and only if the
local field $H_i(\bs) = \sum_j J_{ij} \sigma_j$ has the same sign as $\sigma'_i$, for all $i$.
This condition is then encoded as a product of Heaviside theta functions:
\begin{align} 
\label{eq:indicator}
w (\bs,\bs') = {\prod _{i=1}^{N}} \theta \left( \sigma'_i H_i(\bs) \right).
\end{align}
Now, we are ready to write our starting equation for the number of $L$-cycles.
For given $(L+1)$ spin configurations $\underline{\bs} = \{ \bs(1) , \cdots, \bs (L), \bs (L+1) \}$, the quantity 
$ \prod_{t=1}^{L} w (\bs(t),\bs(t+1))$ now detects the trajectory of a $L$-step evolution of \eqref{eq1}.


The power of this construction comes from the ability that by imposing additional constraints it allows us to select a subset of trajectories. 
In our case, we introduce either the periodic boundary condition $ \bs(L+1) = \bs(1) $ or the skew-periodic boundary condition $ \bs(L+1) = -\bs(1) $, where the minus sign indicates the spin-flipped configuration of $\bs(1)$. 
To handle both cases together, we will use the parameter $P = \pm 1$ to encode the boundary condition  $\bs(L+1) = P \bs(1)$.
With this setting, we define the partition function:
\begin{align}
Z_L^P = \sum_{ \ul{\bs} } \prod_{t=1}^{L} w (\bs(t),\bs(t+1)),
\label{eq:partitionFunctionDef}
\end{align}
where $\ul{\bs}$ now only contains the spin configurations with the boundary condition associated to $P$. 

Certainly, if we impose the periodic boundary condition, the partition function $Z_L^+$ is closely related to the number of $L$-cycles in the system.
However, they are not exactly the same due to the fact that this boundary condition is also satisfied by other cycles of length $L'$ provided that $L'$ is a divisor of $L$, i.e., $L' | L$.
Denoting the number of $L$-cycles by $n_L$, we thus have the following identity
\begin{equation}\label{eq:ZLnL}
Z_L^+ = \sum_{L' | L}  n_{L'} L' \ ,
\end{equation}
where the additional factor $L$ comes from the fact that each $L$-cycle consists of $L$ distinct spin configurations.

Fortunately, we will show that the partition function develops multiple saddle points for each $L'$ satisfying $L' | L$, and thus allows us to choose the desirable saddle point that corresponds to the cycles of length $L$.
For this purpose, it will be convenient to define the two-time overlap parameter $Q(t,s) = \frac{1}{N} \sum_i \sigma_i(t) \sigma_i(s)$ for two different time points $ t<s$.
This measure always falls within the interval $[-1,1]$ and becomes $1$ only when two configurations $\bs(t)$ and $\bs(s)$ are identical.
This implies that the saddle point we are seeking for should be the one satisfying the condition $Q(t,s) <1$ for all pairs of $t < s$.
From now on, by excluding the possibility of having $Q(t,s)=1$ for any two time points $t$ and $s$, we will use the notation $Z_L^+$ to indicate only the contributions for $L$-cycles which cannot be broken into subcycles of smaller length.

There is however another possibility, that follows from the
parity-invariant symmetry imposed by the evolution \cref{eq1}, such that 
$- \sigma_i(t+1) = - \text{sgn} \left[ \sum_{j} J_{ij} \sigma_j(t) \right] = \text{sgn} \left[ \sum_{j} J_{ij} (- \sigma_j(t)) \right]$ 
(see \cite{Bastolla1998} for a more comprehensive discussion).
Namely, if one follows a trajectory that visits the spin-flipped configuration of one of the previously visited configurations at distance $L$ apart, this trajectory automatically forms a cycle of length $2 L$ of the form:
\begin{align} \bs_1 \to \bs_2 \to \cdots 
\to \bs_L \to - \bs_1 \to -\bs_2  \to \cdots \to -\bs_L \to \bs_1. \nonumber
\end{align}
In this case, we say that this trajectory satisfies a skew-periodic boundary condition of length $L$. 
In other words, the trajectory can be broken in two halves, the first going from $\bs_1$ to $-\bs_1$ over a length $L$,
the second going back from $-\bs_1$ to $\bs_1$.
Even though the contribution of this type of trajectories can also be extracted from the periodic boundary 
condition of length $2L$ with the condition $Q(t, t+L) = -1$, we find it more convenient to independently analyze these special trajectories
by considering only the first half of the trajectory, thus introducing the boundary condition $P=-1$ into the partition function.
Combining both contributions for $P= \pm 1$, the overall complexity $\Sigma_L$ for each $ L $ is then given by
\begin{align}
\label{eq:overallComplexity}
\Sigma_L = 	\left\{\begin{array}{lr}
\Sigma_L^+ & \text{for  odd } L \ , \\
\max ( \Sigma_L^+ , \Sigma_{L/2}^-), & \text{for even } L \ .
\end{array}\right.
\end{align}
In the next section we show how to compute $Z_L^P$ and from it extract $\Si_L^P$.

\subsection{Average of the number of $L$-cycles}
In this section, we present a detailed analysis of the annealed average of $Z_L^P$ over different realizations of $\{J_{ij}\}$.
As usually done in fully-connected models, our aim is to transform $\QuenchedAverage{Z_L^P}$ into an integral over a set of variables $\bf X$ and in turn extract the asymptotic behavior via the saddle point method:
\begin{align} \label{eq:partitionFunctionAverage}
\QuenchedAverage{Z_L^P} 
=  \sum_{\bf X^*}  A_L({\bf X^* }) e^{N \Sigma_L^P (\bf X^*) },
\end{align}
with corrections of order $O(N^{-1})$.
Here, the starred variables $ {\bf X^* } $ indicate the extrema of $\Sigma_L^P({\bf X})$.
Once the form of the partition function \cref{eq:partitionFunctionAverage} is determined, the typical cycle length in the system is given by the one yielding the largest exponential growth rate $\Sigma_L^P$, provided that $N$ is sufficiently large.

After some calculations, detailed in \cref{appendix:EvalActionL}, 
we show that 
$\QuenchedAverage{Z_L^P}$ can indeed be cast into a saddle point form over two symmetric matrices of variables $Q(t,s), R(t,s)$ and one non-symmetric matrix $S(t,s)$.
Namely, $\QuenchedAverage{Z_L^P}$ reads
\begin{align}
\QuenchedAverage{Z_L^P}   \sim \int_{R, Q, S} e^{N \Sigma_L^P }
\end{align}
up to a multiplicative constant.
The \textit{complexity} is then given by
\begin{align}
\Sigma_L^P &=
- \sum_{t > s} 
R(t,s) Q(t,s) - \frac{\eta}{2} \sum_{t,s}
S(t,s) S(s,t) + \log
\left( \mathcal{Z}_L^P
\right),
\label{ComplexityOriginal}
\end{align}
with one site partition function
\begin{align}
\mathcal{Z}_L^P & = 
\MeasureSite{P}
e^{
	-\frac{1}{2} \sum_t \lambda(t)^2} 
e^{	- \sum_{t>s} Q(t,s) \l(t) \s(t+1) \l(s) \s(s+1) + \sum_{t>s} R(t,s) \s(t) \s(s) }  e^{  \sum_{t,s} \eta S(t,s)  I\l(t) \s(t+1) \s(s) 
},
\end{align}
where $ \MeasureSite{P} (...) $ refers to the (weighted) integrations and the sums over possible values of $\lambda(t)$ and $\sigma(t)$ 
given by
\begin{align}
\MeasureSite{P}(...) = \sum_{ \SumConstraintOneSite{P} }  \int \prod_t \frac{d\lambda(t)}{2\pi (I \lambda(t) + \e)} (...),
\end{align}
with $\e$ being a positive infinitesimal number, and $I$ is the imaginary unit.

For each $L$, this expression should be extremized with respect to $Q(t,s)$, $R(t,s)$ and $S(t,s)$.
As it can be checked, the parameter $\epsilon$ only enters in $\Sigma_L^P$ via the single parameter $\eta$ defined in \cref{eq3}.
In principle, the multiplicative prefactor $A_L^P$ can be computed within this framework by computing the corrections to the saddle point, 
and some specific values of $A_L^P$ will be reported for $\epsilon=1$, where $\Sigma_L^P$ vanishes.
Note that from Eq.~\eqref{eq:ZLnL} and the following discussion we obtain $n_L \sim (A_L/L) e^{N\Sigma_L}$.

For the periodic boundary condition, one can make a further simplification. 
Because cycles are by definition symmetric  under translation of time (i.e. $t\to t+t_0$ for any $t_0$), one may employ an ansatz that $ Q(t,s) $ and $ R(t,s) $ are only functions of the distance $|t-s|$, i.e. write $ Q(|t-s|) $ and $R(|t-s|)$, respectively.
It should be noticed that the distance $|t-s|$ is defined by taking into account periodic boundary condition, i.e. by considering the minimum difference 
between $t$ and all the periodic images of $s$.
Similarly, one can write $S(t,s) = S(s-t)$.
One crucial difference, in this case, is that the function $S(s-t)$ is not even in $s-t$, as the two time directions
are not equivalent.

\section{Results}
\label{sec:result}

In the following, we present the results for the average number of cycles of given length $L$, i.e. $n_L$, of the form 
\begin{equation}\label{eq:nLscal}
n_L (\epsilon,N) = \frac{A_L (\epsilon)}{L} \; e^{\Sigma_L(\epsilon)  N} \ .
\end{equation} 
As discussed above, the form of the prefactor with $L$ at the denominator is a consequence of Eq.~\eqref{eq:ZLnL}.
Note that
 $\Sigma_L$ is independent of the choice of distribution as long as the distribution is symmetric with a finite second moment, whereas $A_L$, which is independent of $N$, depends also on the fourth cumulant of distribution. 
One may even generalize this result to non-symmetric distributions with zero mean without changing $\Sigma_L$.
Since the behavior is mainly determined by $\Sigma_L$ for sufficiently large $N$, we first focus on determining $\Sigma_L$.

Computing $\Sigma_L$ for arbitrary $L$ involves finding saddle points of three matrices $Q(t,s)$, $R(t,s)$ and $S(t,s)$, which is a non-trivial problem.
We thus first focus on determining the behavior of $\Sigma_L$ for $L=1,2,3$ by numerically optimizing for the above matrix elements. 
To do this, one needs to consider  $\Sigma_L^+$ for $L=1,2,3$ and $\Sigma_L^{-}$ for $L=1$ as suggested by \cref{eq:overallComplexity}.
We then study in detail the case $\epsilon=1$ for arbitrary $L$, giving insight into the numerically observed phase transition. 
Surprisingly, we will show that, at $\epsilon=1$, $\Sigma_L =0$ for all $L$, thus the behavior of $A_L$ plays a crucial role to determine the 
relative importance of cycles of length $L$.

In the following, the calculation will be performed in terms of $\epsilon$ or $\eta$, depending on the specific convenience.

\subsection{Complexity of fixed points: $L=1$}

\begin{figure*}
	\includegraphics[width=.95\textwidth]{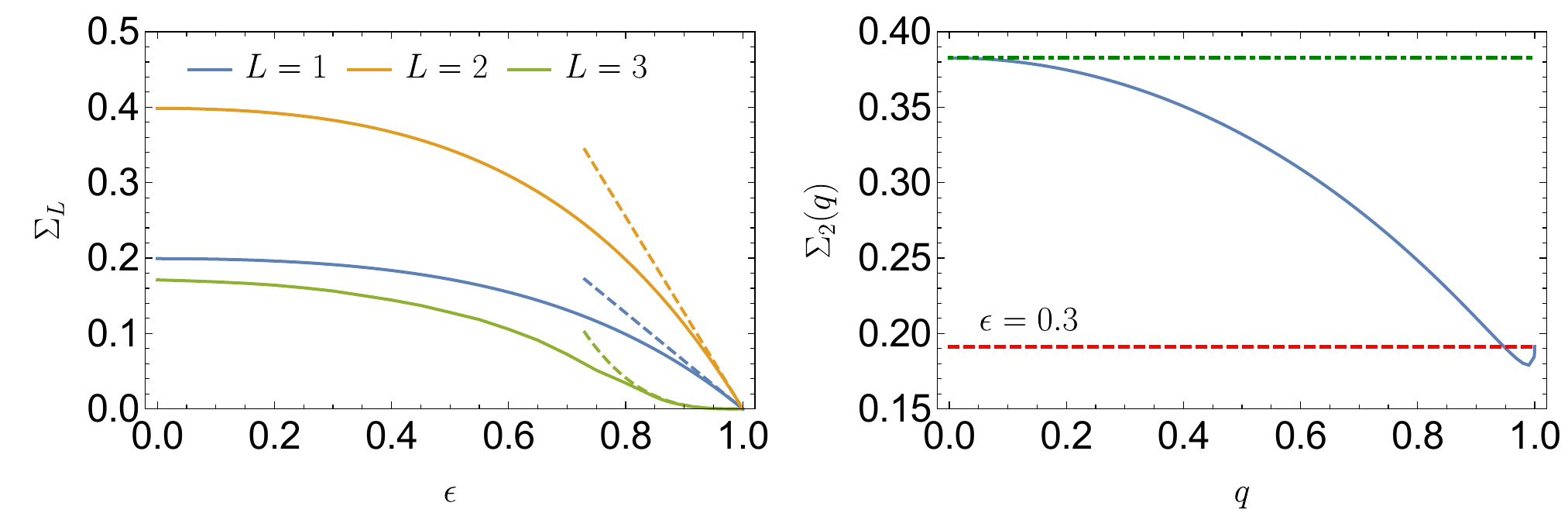}
	\caption{(Left) The function $\Si_L(\ee)$ as a function of $\ee$ for $L=1,2,3$. 
Dashed lines indicate the lower-order contribution computed from the perturbation theory in $\h \sim 1-\ee$. The curve for $L=2$ is always larger than the other curves for all $\epsilon <1$. (Right) The constrained $\Si_2(q)$ for different values of $q$ at $\epsilon=0.3$.
The two local maxima at $q=0$ and $q=1$ correspond to $\Sigma_2$(dot-dashed) and $\Sigma_1$(dashed), respectively. 
		This implies that the saddle point corresponding to $q=0$ gives the dominating contribution.\label{fig:SigmaTheory}
	}
\end{figure*}

In this special case the only nontrivial parameter is $S \equiv S(t,s) = S(1,1)$.
Namely, the complexity $\Sigma_1^P$ reads
\begin{align}\nonumber
\overline{Z_1^P} &\sim \int dS
e^{-\frac{1}{2} N  
	\eta
	S^2
}
\left(
\sum_{ \s } \int \frac{d\l}{2\pi (I\l + \e)} 
e^{
	\frac{1}2 [ - \l^2  + 2 \eta P S  I\l ]
}
\right)^N \sim \int dS e^{N \Sigma_1^P } \ ,
\end{align} 
where 
\begin{align}
\label{eq:Complexity1}
 \Sigma_1^P & = -\frac{1}2 \eta S^2  + \log
\left(
2 \int \frac{d\l}{2\pi (I\l + \e)} 
e^{
	\frac{1}2 [ - \l^2  + 2 \eta P S  I\l ]
}
\right) \nonumber \\
&= -\frac{1}2 \eta S^2 + \log 2  + \log \Phi(\eta P  S),
\end{align}
where $\Phi(x)$ is the CDF of the standard Gaussian distribution, i.e., $ \Phi(x) = \frac 1 {\sqrt{2\pi}} \int_{-\infty}^x e^{-t^2/2} \, dt $.
The fact that the substitution $S \to P S$ removes the occurrence of $P$ implies 
$\Sigma_1^+ = \Sigma_1^-$ and furthermore  $\overline{Z_1^+} = \overline{Z_1^-}$.
For $\ee=0$ we find back the Tanaka-Edwards result $\Sigma_1 = 0.19923\ldots$. 
At $\ee=1$, we find $\Si_1 = 0$. The whole $\Sigma_1(\epsilon)$ vs. $\epsilon$ is reported in figure \ref{fig:SigmaTheory} (left panel) as full blue line.

\subsection{Complexity of limit cycles with $L=2$}

For $L=2$, the non-trivial parameters are the following:
$
Q_d \equiv Q(1,2), R_d \equiv R(1,2),$
$	 S_1 \equiv S(1,1), S_2 \equiv S(2,2), S_{12} \equiv S(1,2)$
and $ S_{21} \equiv S(2,1)
$, which already describes the high-dimensional nature of our problem.
The complexity \cref{ComplexityOriginal}
in this case reads
\begin{align}
& \Si_2^P = -  R_d Q_d - \frac{1}{2} \eta (S_1^2 +  S_2^2 + 2 S_{12} S_{21}) \\ & + \log
\Bigg\{
2 \sum_\s e^{ R_d \s} \int \frac{d\l_1 d\l_2}{4\pi^2 (I\l_1 + \e)^2} 
e^{-\frac{1}2 (\l_1^2 + \l_2^2 +2  Q_d P \s \l_1 \l_2 ) } e^{ \eta [ I \l_1 (S_1 \s + S_{12}) + I\l_2 (S_1\s + P S_{21}) ]
}
\Bigg\}. \nonumber
\end{align}
First, we show that there exists a saddle point that satisfies the following conditions: $Q_d = R_d = S_1 = S_2 =0$.
Note that these terms only appear with $\sigma$ as the argument of the logarithm.
Because of this, the derivative of the logarithmic term with respect to any of these variables yields an additional $\sigma$, and subsequently cancel out when summing over $\sigma$, which verifies the saddle point condition at the above conditions.
Under these conditions, collecting the remaining terms, we find a result similar to the one for $S_1$ in~\cref{eq:Complexity1}:
\begin{align}
\Si_2^P 
&= 	- P  \eta S_{12} S_{21}
+ 2 \log 2 
+ \log \Phi(\eta S_{12}) 
+ \log \Phi(\eta S_{21}),
\end{align}
where we have applied a transformation $ S_{12} \to P S_{12}$ which makes the parameter space symmetric under the exchange $S_{12} \leftrightarrow S_{21}$.
If this symmetry is unbroken, the saddle point should satisfy $S_{12} = S_{21} \equiv S$. 
Within this ansatz, the complexity is further simplified to 
\begin{align}
\Si_2^P &= 	- \eta P S^2
+ 2 \log 2 
+ 2\log \Phi(\eta S).
\end{align}
Surprisingly, this implies 
\begin{equation} 
\fbox{$ \displaystyle
	\Sigma_2^+(\eta) = 2 \Sigma_1^+(\eta)
	$  ,}
\label{eq:Solution2}
\end{equation} 
and
\begin{equation} 
\fbox{$ \displaystyle
	\Sigma_2^-(\eta) = \Sigma_2^+(-\eta) = 2 \Sigma_1^+(-\eta)
	$.}
\label{RelationSigma2}
\end{equation} 
This is the first important result of the present paper. To our knowledge, this is the first derivation of the number of cycles of length larger than one.

As shown in \cref{fig:SigmaTheory} (left), the exponential growth rate $\Sigma_2^+(\eta)$ is always positive in the range $0 < \eta \le1$.
This automatically means from \cref{RelationSigma2} that $\Sigma_2^-(\eta)$ is positive for $\eta <0$.
Thus, according to Eq.~\eqref{eq:overallComplexity}, 
the skew-periodic trajectories of length $4$ 
give a positive contribution to $\Sigma_4(\eta)$ when $\eta<0$.
Furthermore, it is worth noting that $Q_d = 0$ suggests that two configurations comprising 2-cycles each are spatially uncorrelated.
This result provides a  solid ground for the annealed approximation that is used in \cite{Bastolla1997} for the case of $L=2$.

To check whether this solution yields the dominating contribution, we should further determine whether there are other saddle points.
To provide some evidence, we consider the sub-problem of fixing $Q_d$ to have a prescribed value $q$, and optimizing only over the other variables. 
If there are other solutions, the complexity $\Sigma^P_2(q)$ should develop different local maxima.
In \cref{fig:SigmaTheory}, we numerically confirm that $q =0$ is indeed the solution for the case $P=1$. 
The figure shows that there is another solution at $q =1$ as well, which, as expected, corresponds to the solution of 1-cycle (i.e., $\Sigma_1^+$).

\subsection{Complexity of limit cycles for $L=3$ with $P=1$}

Here, we repeat the same procedure for $L=3$. 
To reduce the unnecessary complexity, we focus our interest to the case $P=1$.
In this particular case, we have five non-trivial parameters, namely
$ Q_1, R_1, S_{-1}, S_0$ and $ S_1 $ where the argument indicates the difference between two time points.
Then, the complexity reads
\begin{align}
\Si_3^+ = -3 Q_1 R_1-\frac{1}{2} \eta \left(3 S_0^2+6 S_1 S_2\right) +  \log
\left( \mathcal{Z}_3^+
\right),
\label{eq:Solution3}
\end{align}
where 
\begin{align}
\mathcal{Z}_3^+ & = 
\MeasureSite{+}
e^{
	-\frac{1}{2} \sum_t \lambda(t)^2} 
e^{	-Q_1 \sum_{t>s} \l(t) \s(t+1) \l(s) \s(s+1)} e^{	R_1 \sum_{t>s}  \s(t) \s(s) }  e^{  \sum_{t,s} \eta S(t,s)  I\l(t) \s(t+1) \s(s) \nonumber
}.
\end{align}
Obviously, the numerically challenging part to determine the saddle point is the evaluation of $\mathcal{Z}_3^+$, which is a three-dimensional complex-valued integral. 
Instead of performing a direct integration, we can convert this problem to a problem of finding an expectation value of the Gaussian measure by employing a Hubbard-Stratonovich transformation, which yields
\begin{align}
\mathcal{Z}_3^+ & =  \mathrm{E}_z
\sum_{ \SumConstraintOneSite{+} }  e^{	R_1 \sum_{t>s}  \s(t) \s(s) } 
\prod_{t=1}^{3} \Phi\Parenthesis{
	\frac{\s(t) (z \sqrt{Q_2} + S_{-1}\sigma(t-1) + S_1 \s(t) + S_{1} \s(t+1) ) }{\sqrt{1-Q_2}}
},
\end{align}
where $ \mathrm{E}_z $ refers to the average with respect to the standard Gaussian variable $z$.
After this conversion, we arrive at a real-valued one dimensional Gaussian integral which is numerically much more feasible.
Figure \ref{fig:SigmaTheory} shows a plot of $\Sigma_3$ (which is equal to $\Sigma_3^+$) numerically optimized over five variables.
The fact that $\Sigma_3$ always lies below $\Sigma_2$ implies that the 3-cycles are exponentially outnumbered by the cycles with length two for the entire range of $0 \le \epsilon \le 1$. 

\begin{equation} 
\fbox{$ \displaystyle
	\Sigma_3(\epsilon)  \le \Sigma_2(\epsilon) 
		\; \; \; \; \; \; $ for $0 \le \epsilon \le 1$.}
\label{RelationSigma3}
\end{equation} 

%

\subsection{Vanishing of the complexity of limit cycles for arbitrary $L$ at $\epsilon=1$}

As seen from the previous case, performing a saddle point calculation becomes quickly unmanageable as we increase $L$.
However, we have already captured one important observation: for arbitrary $\epsilon<1$, we have $ \Sigma_2 > \Sigma_L$ for $L=1, 3$.
Surprisingly, our numerical studies suggest that this behavior is robust also for larger $L$'s. 

For small enough $\epsilon \lesssim 0.7$, corresponding to $\eta \gtrsim 0.5$ according to Eq.~\eqref{eq3}, 
the typical length of limit cycles is two, whereas longer cycles come into play more frequently as $ \epsilon $ increases.
To understand this behavior, let us focus on $\epsilon=1$, which corresponds to $\eta=0$, i.e. to fully asymmetric coupling matrices. 
In this case, the matrix $S(t)$ does not appear in the complexity, see Eq.~\eqref{ComplexityOriginal},
and we can focus only on the two matrices $Q(t)$ and $R(t)$.
Further, we note that at $\epsilon=1$, the choice $R(t) =0$ and $Q(t) =0$ (for $t\neq 0$) verifies the saddle point equations, 
for all finite values of $L$, and the corresponding value of complexity is $\Sigma_L^P= 0$.
We will conjecture that this is the dominant saddle point at $\epsilon=1$, and therefore all the complexities vanish in this case;
this is consistent with the solutions for $L \le 3$ we have obtained from the previous analysis.
It also implies that any pairs of two configurations comprising a $L$-cycle is uncorrelated, which is once again
consistent with the annealed approximation adopted in \cite{Bastolla1997} for the case $\epsilon=1$.

\begin{equation} 
\fbox{$ \displaystyle
	\Sigma_L(\epsilon$=1$)  =0
		\; \; \; \; \; \; $ for $L=1, 2, ... \infty$.}
\label{RelationSigmaL}
\end{equation}

\begin{figure*}
	\begin{center}
	\includegraphics[width=.95\textwidth]{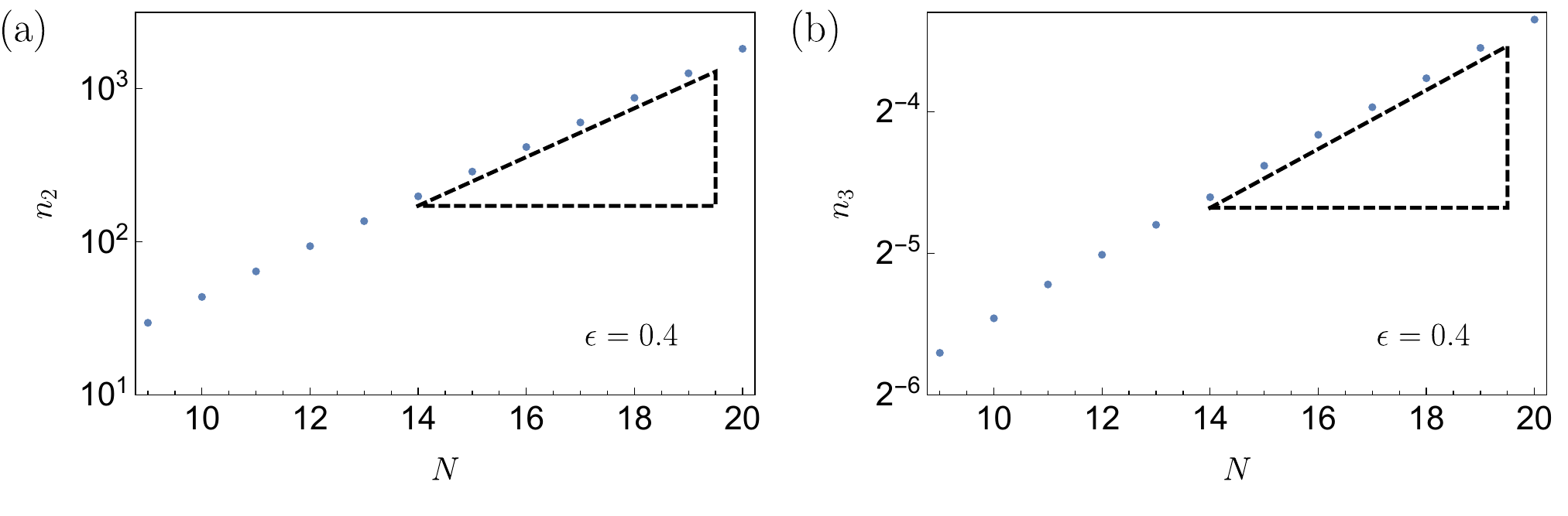}
	\caption{Numerical simulations of $n_L$ for $L=2$ (a) and $L=3$ (b) at $\epsilon= 0.4$;
		The slope of dashed triangles indicates the exponential growth rate of $n_L$, i.e., $\Sigma_L$ computed using \cref{eq:Solution2,eq:Solution3}, respectively.
		Even for relatively small system size $N < 20$, the growth of $n_L$ is well described by its asymptotic growth rate $\Sigma_L$ in the entire range of~$\epsilon$.
		}
		\end{center}
	\label{fig:Simulation}
\end{figure*}

\subsection{Complexity of limit cycles for arbitrary $L$ close to $\epsilon=1$}
\label{sec:AAAAAA}

Given the simple structure of the solution $R(t)$ and $Q(t)$ at $\epsilon=1$, 
	we can analyze $\Sigma_L$ perturbatively as a power-series of $\eta$, i.e. perturbatively for $\epsilon$ close to $1$.
Relegating the detailed steps to \cref{appendix:Perturbation}, we simply summarize the leading behaviors of $\Sigma_L^P$:
\begin{enumerate}
	\item first order :
	\begin{align}
	\Sigma_1^P &= \frac{1}{\pi} \eta + \text{h.o} \ , \quad
	\Sigma_2^P = \frac{2P}{\pi} \eta + \text{h.o}  \ ,
	\end{align}
	\item second order
	\begin{align}
	\Sigma_4^P &= \frac{8}{\pi^2} \eta^2 + \text{h.o}\ , \quad
	\Sigma_L^- = O(\eta^3) \,\,\,\, \text{for all other $L$} \ ,	
	\end{align}	
	\item third order
	\begin{align}
	\Sigma_3^+ &= \frac{20}{\pi^3} \eta^3 + \text{h.o} \ , \quad 
	\Sigma_6^+ = \frac{40}{\pi^3} \eta^3 + \text{h.o}\ , \nonumber\\	
	\end{align}
	\item fourth order
	\begin{align}
	\Sigma_8^+ &= \frac{224}{\pi ^4} \eta^4 + \text{h.o} \ , \quad 
	\Sigma_L^+ = O(\eta^5) \,\,\,\, \text{for all other $L$}	\ ,
	\end{align}	
\end{enumerate}
Although this analysis is not easily extended to higher orders, it uncovers two interesting behaviors. 
First, for odd $L$, we found 
\begin{center}
\fbox{$2 \Sigma_L^+(\eta) =  \Sigma_{2L}^+(\eta)
\;\;\;\;$ for small $\eta$\ .
}
\end{center}
which generalizes Eq. (\ref{eq:Solution2}) to all $L$. Second, we find that for even $L$: 
\begin{center}
\fbox{$\Sigma_L^+(\eta)= \eta^{L/2} W_L \;\;\;\;$; $\;\;\;\; W_L = \frac{2^{L/2} L!}{\left(\frac{L}{2}+1\right)! \frac{L}{2}! \pi^{L/2}}$ \ .
}
\end{center}
Assuming that this trend continues to be satisfied, the asymptotic behavior for $W_L$ reads
 $W_L \sim e^{(3 \log 2 - \log \pi ) \frac{L}{2}}$, hence
 \begin{equation}\label{eq:SiLscal}
\Sigma_L^+ \sim \begin{cases} e^{ \kappa  \frac{L}{2}} & \text{for even $L$} \ , \\
e^{ \kappa  L} & \text{for odd $L$} \ , 
\end{cases}\qquad
\text{with} \qquad
\kappa = \log( 8\h/\pi) \ .
\end{equation}
Within this conjecture, one can conclude that $\Sigma_L$ decreases exponentially with $L$ for small enough $\eta$,
where $\k<0$.
Note that $\k$ becomes positive for $\eta>\pi/8\approx0.393$, but for such large $\eta$ the perturbation theory developed 
in this section is certainly not correct (also because an exponential growth of $\Sigma^+_L$ with $L$ would be incompatible with
the bound $\Sigma^+_L < \log(2)$ which follows from the fact that the total number of neuron states is $2^N$).

\subsection{Distribution of limit cycles of length  $L$ at $\epsilon=1$}

Having derived that $\Sigma_L^P = 0$ for $\ee=1$ and all finite lengths $L$, it is important to understand how the prefactor  grows with $L$, i.e., $\PartitionFunc{L} = A_L^P \exp(N \Sigma_L^P) = A_L^P$.
In \cref{appendix:EvalPrefactor}, we explicitly computed 
\begin{align}
A_L^P = M_P(L) \exp(\mathcal{H}) \ ,
\end{align}
where
\begin{align}
\mathcal{H} = -P \frac{2}{\pi} \delta_{L,2} + 
\mathcal{A} \frac{6}{\pi^2} \SquareBracket{
	\delta_{L,2} + 2 P \delta_{L,4}
} \ ,
\label{ExponentialCorrectionAtOne}
\end{align}
$M_P(L)$ is a constant converging to one exponentially fast upon increasing $L$, 
and $\mathcal{H}$ is a distribution-dependent constant which is non-zero only for $L=2$ and $L=4$ (see \cref{appendix:EvalPrefactor} for details).
Consequently, for sufficiently large $L$, we have $ \PartitionFunc{L} \simeq 1 $.
However, this might seem surprising, because $L$ cannot be arbitrarily large in finite systems. 
Certainly this is because of the limit $N \to \infty$ at a fixed $L$. 
Thus for finite $N$, there must be a cut-off function $ L_c(N) $ which effectively determines the maximum cycle length.
The obvious upper bound is $L_c(N) \le 2^N$. 

\begin{figure*}
\begin{center}
\includegraphics[width=.85\textwidth]{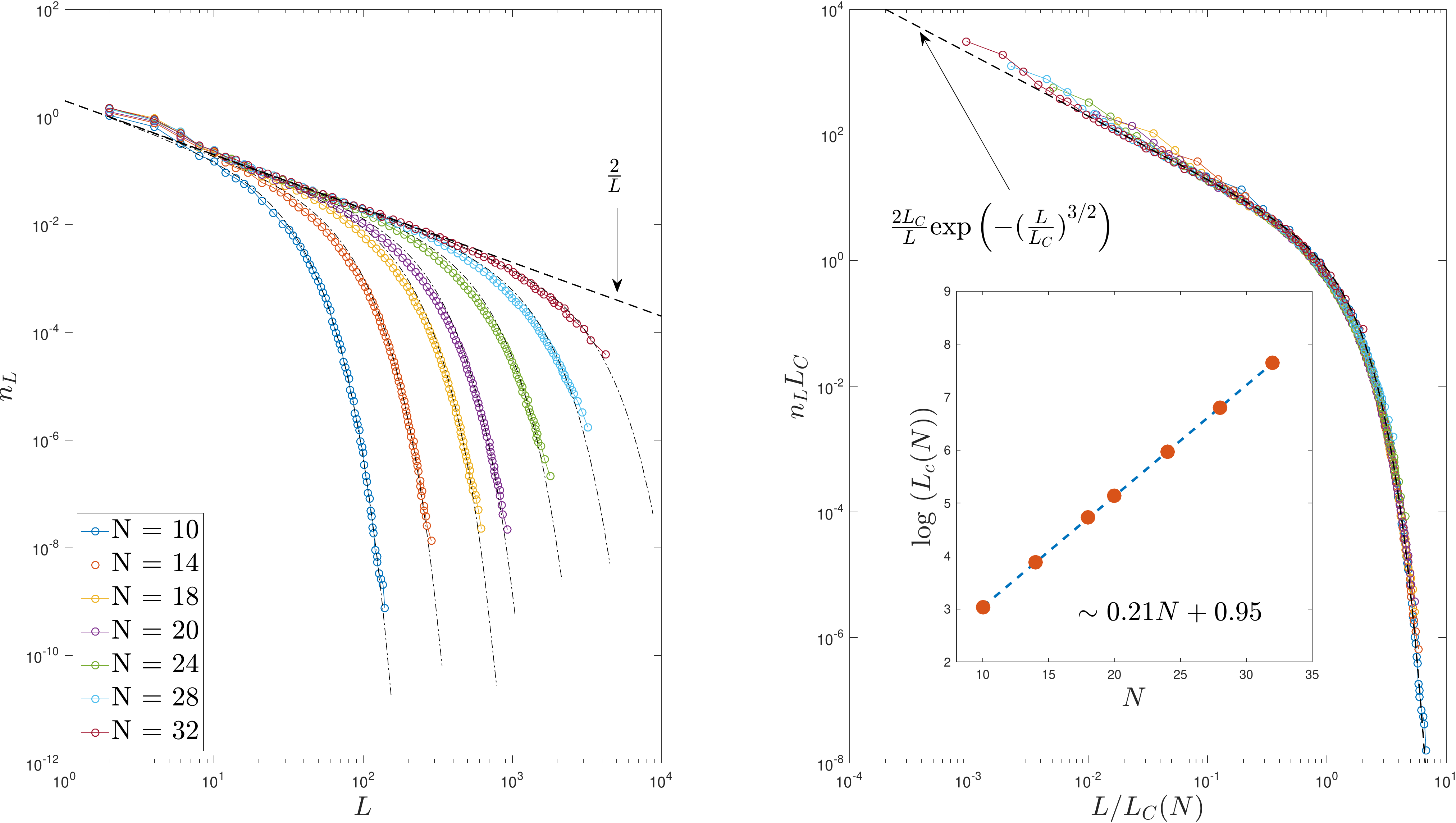}
\end{center}
\caption{
(Left)
Numerical estimation of the average number of limit cycles $n_L$ as a function of the cycle length $L$, for binary couplings, with 
various network sizes $ N $ at $\epsilon$ = 1. 
The data are well fitted by the form $n_L = (2/L) \exp\{-[L/L_c(N)]^{3/2}\}$ (dot-dashed lines).
(Right) Using the fitted value of $L_c(N)$ all curves can be collapsed. (Inset) The values of $L_c(N)$ are well fitted by
$L_c(N) \sim \exp[0.95+0.21 N]$.
}
  \label{simulazioni4}
\end{figure*}

\begin{figure}
\begin{center}
\includegraphics[width=.85\textwidth]{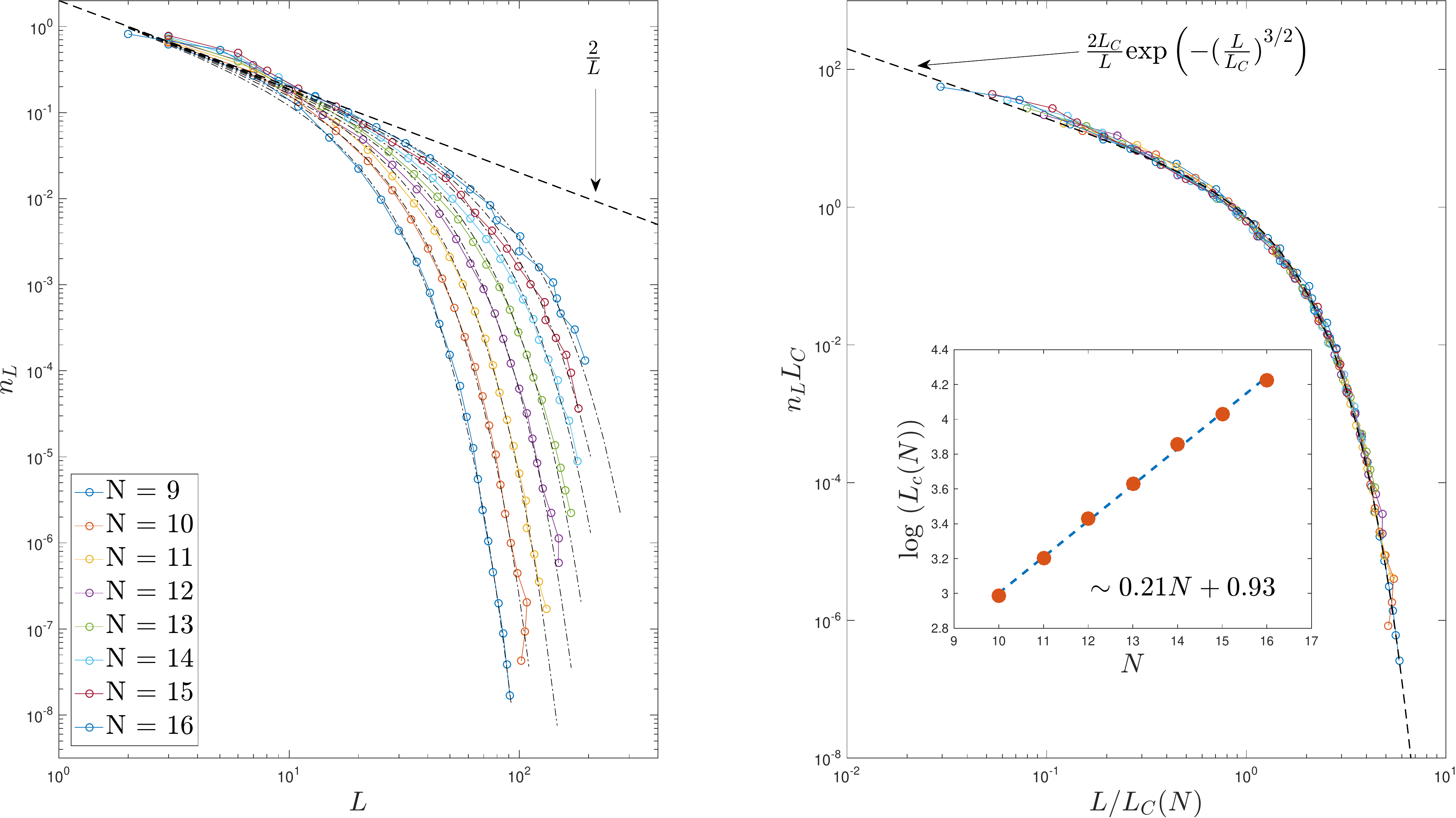}
\end{center}
\caption{
(Left)
Numerical estimation of the average number of limit cycles $n_L$ as a function of the cycle length $L$, for Gaussian couplings, with 
various network sizes $ N $ at $\epsilon$ = 1. 
The data are well fitted by the form $n_L = (2/L) \exp\{-[L/L_c(N)]^{3/2}\}$ (dot-dashed lines).
(Right) Using the fitted value of $L_c(N)$ all curves can be collapsed. (Inset) The values of $L_c(N)$ are well fitted by
$L_c(N) \sim \exp[0.95+0.21 N]$.
}
  \label{simulazioni5}
\end{figure}

To better clarify this point, in the left panels of figures \ref{simulazioni4} (binary distribution of $J$) and \ref{simulazioni5} (Gaussian distribution) we report the distribution function $n_L = \overline{n_L(\mathbf J)}$ as a function of the cycle length $L$, for various network sizes $N$ at $\epsilon$ = 1. Only even $L$ are reported. The graph confirms the power law shape of $n_L$ at small $L$ and the existence of a cut-off at larger $L$, which shifts towards larger $L$ with increasing network size. 
The $N$ dependence of the cut-off $L_c(N)$, as defined by the condition $n_{L_c}$=$constant$, with a constant in the range $10^{-4} $-$10^{-5}$ , confirms the exponential dependence on $ N $ and reveals that the logarithmic slope $\alpha$  is  about 0.21 in the whole range ($\alpha$=0.21 $\pm$ 0.01).

Moreover, we observe that, at $\epsilon$=1, $n_L$ is well represented by the following function for all values of $N$ and both $J$ distributions (see the dot-dashed lines in the right panel of figures \ref{simulazioni4} and \ref{simulazioni5}): 

\begin{equation} 
\label{func}
\fbox{ $ \displaystyle
n_L  = \frac{2}{L} \exp \left( - \left[ \frac{L}{L_c(N)} \right]^{3/2} \right)
\;\;\;\;$;$ \displaystyle \;\;\;\; L_c(N)=\exp \left ( \alpha N + \beta \right )$}
\end{equation}
for even $L$.

Incidentally, this result indicates that the quantity $n_L(\epsilon$=1$,N)$ depends only on a scaling parameter $L_c(N)$, as demonstrated by the right panel of figures \ref{simulazioni4} and \ref{simulazioni5} where $L_c(N) n_L$ is reported as a function of $L/L_c(N)$ and all data points collapse on a single curve given by $x^{-1} e^{-x^{3/2}}$.

Within the annealed approximation, it has been shown that the cut-off $L_c^{\mathrm{ann}}(N)$ is given by the same form with $\alpha \simeq 0.228$ \cite{Bastolla1997}.
Despite the fact that this result was obtained from the mean number of cycles weighted by the size of basins, this striking similarity suggests a marginal role of basin weights in determining the cutoff.

\subsection{Distribution of number of cycles within one sample}
\begin{figure*}
	\begin{center}
		\includegraphics[width=.85\textwidth]{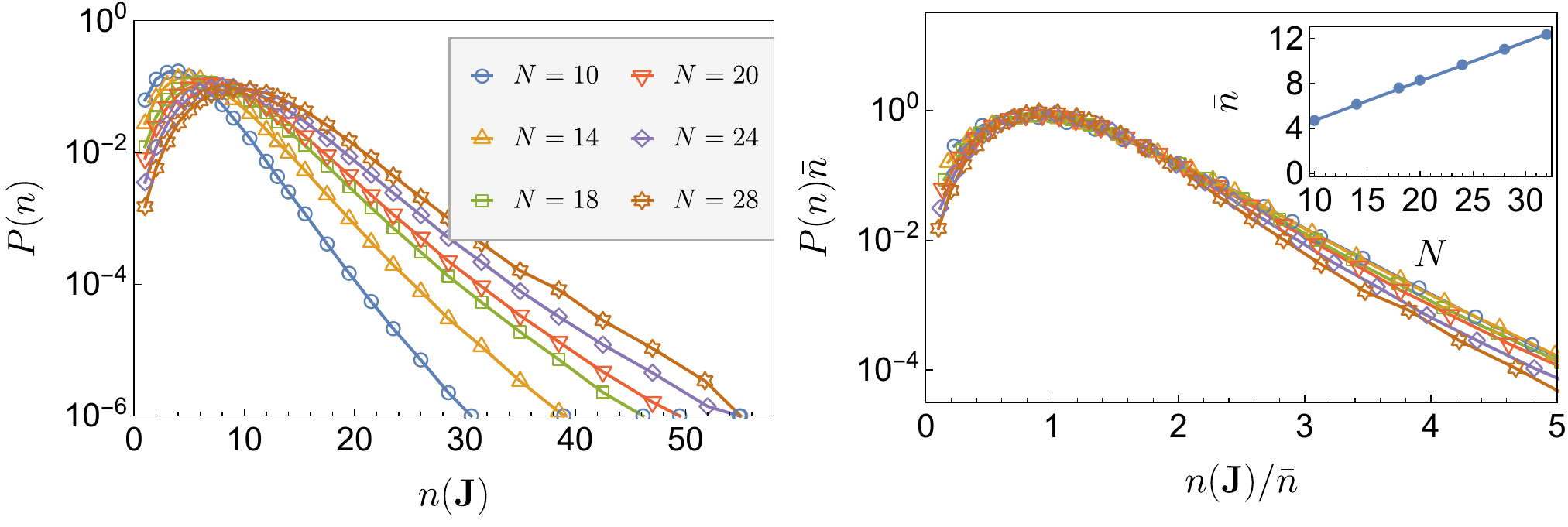}
	\end{center}
	\caption{
		(Left)
		Histogram of the number of cycles $n (\mathbf J)$ within a sample for binary couplings at $\epsilon = 1$.	
		The position of the peaks moves to the right with increasing network size $N$.
		(Right) The scaled distribution of the number of cycles $n(\mathbf{J})$ by the mean value $\bar n$. 
		From our scaling theory \cref{func,PositionPeak}, the peak should be collapsed when scaled by $\bar n$ given in \cref{PositionPeak}.
		The distribution seems to have a stretched exponential tail, and thus decays slower than a Poisson distribution.
		(inset) Plot of $\bar n$ as a function of $N$. It shows a clear linear relationship.
	}
	\label{NumberOfCycle}
\end{figure*}

This scaling equation \cref{func} provides valuable information of the number of cycles $\bar n$ within one sample for the case $\epsilon=1$.
This quantity $\bar{n}$ is simply given by 
\begin{align}
	\bar n = \sum_L n_L .
\end{align}
Using the scaling form in \cref{func} and also taking into account the contribution of cycles of odd length (see Appendix E), we establish a linear relationship between $\bar n $ and $ N$:
\begin{align}
\bar n \simeq 0.35 N + 1.2,
\label{PositionPeak}
\end{align}
where the coefficients were estimated from the simulation results for $n_L$ (See the inset of \cref{NumberOfCycle} (Right)).
In contrast to the exponential growth of $L_c(N)$, this implies that the mean number of cycles behaves rather mildly, and thus does not lead to proliferation of many cycles.

To check the consistency, we study the distribution of number of cycles $n(\mathbf{J})$. 
According to our estimate \cref{PositionPeak}, the mean value of the distribution increases roughly proportional to $N$.
In \cref{NumberOfCycle} (Left), our statistics are shown to be moderately distributed and their characteristic size is well described by the peak of distribution.
Furthermore, it is clearly observed that the peaks are moving to the right with the network size $N$.
Given this fact, the region around the peak should be well collapsed according to our theory \cref{PositionPeak} (See \cref{NumberOfCycle} (Right)).

\subsection{Average cycle length}

Now, we present an argument to estimate the divergence point of
the average length of cycles $\bar L$.
First, we note that from Eq.~\eqref{eq:nLscal} we have
\begin{equation} \label{eq:barL}
\bar L  = \frac{A_1 e^{N\Sigma_1} + A_2 e^{N \Sigma_2} +\sum_{L=3}^{L_c(N)} A_L e^{N \Sigma_L}  }{A_1 e^{N\Sigma_1} +\frac{A_2}{2} e^{N \Sigma_2} +
 \sum_{L=3}^{L_c(N)} \frac{A_L}{L} e^{N \Sigma_L} } 
\sim \frac{2 + \frac{1}{A_2} \sum_{L=3}^{L_c(N)} A_L e^{N (\Sigma_L-\Sigma_2)}  }{1 + \frac{2}{A_2} \sum_{L=3}^{L_c(N)}  \frac{A_L}{L} e^{N (\Sigma_L-\Sigma_2)} }  \ ,
\end{equation}
where we neglected the term $L=1$ because we already know that $\Sigma_1<\Sigma_2$ for all $\ee<1$.

We showed in the previous sections that at $\ee=1$ we have $A_L\approx 1$, and we are going to assume that this result holds
in the vicinity of $\ee=1$ as well.
Moreover, for $\ee\sim 1$ we also have from Eq.~\eqref{eq:SiLscal} that $\Sigma_L$ decreases exponentially with $L$, leading to the conjecture
that $\Sigma_2$ is the largest complexity.
Because $A_L\approx 1$ and $\Sigma_L\approx 0$ at large $L$, the sums in Eq.~\eqref{eq:barL} have leading terms that behave as
\begin{equation}
\sum_{L=3}^{L_c(N)} A_L e^{N (\Sigma_L-\Sigma_2)}  \approx L_c(N) e^{-N \Sigma_2} \ ,
\qquad
\sum_{L=3}^{L_c(N)} \frac{A_L}{L} e^{N (\Sigma_L-\Sigma_2)}  \approx \log[L_c(N)] e^{-N \Sigma_2} \ .
\end{equation}
Interestingly, if we assume that $L_c(N,\epsilon) \approx L_c(N,\epsilon =1)\approx e^{\a N}$ 
 increases exponentially fast as discussed above,
the second sum is always exponentially small in $N$, while the first sum can either vanish or diverge with $N$, 
leading, at leading order in $N$, to
\begin{equation} 
\label{eq:barL2}
\fbox{ $ \displaystyle
\bar L -2  \approx \frac{1}{A_2} L_c(N,\ee) e^{-N \Sigma_2(\ee)} \approx  e^{N [\a - \Sigma_2(\ee)]} 
$}
\end{equation}
This formula implies a transition from 
$\bar L  =2$ when $\a < \Sigma_2(\epsilon)$ to $\bar L \to \infty$ when $\a > \Sigma_2(\epsilon)$, 
thus confirming our numerical results for the transition in average cycle lengths \cite{Nutzel1991,Bastolla1998}.

The condition $\Sigma_2(\epsilon_c)=\alpha$ identifies the critical value of the parameter $\epsilon$ where the transition to chaos takes place. 
This condition gives $\epsilon_c=0.797$ or $\h_c\approx 0.393$.
Note that curiously, this critical value coincides with the point where Eq.~\eqref{eq:SiLscal} indicates an unphysical exponentially growing $\Sigma_L$, suggesting a breakdown of perturbation theory.

Even though this analysis provides a strong evidence for the transition to chaos, the hypothesis that our cutoff $L_c$ does not change as a function of $\epsilon$ has to be challenged. 
In fact, a refined analysis indicates that this hypothesis is actually not reliable (See \cref{appendix:MoreOnCycleLength}). 
Nevertheless, our main findings remain correct apart from the precise position of $\epsilon_c$, that is estimated to be slightly larger. Specifically, we found $\epsilon_c =0.835$ ($ \eta_c = 0.321 $).

Our estimate of $ \h_c $ turns out to be smaller in comparison to the value $0.5$ obtained from the average weighted  by the size of the basins of attraction \cite{Bastolla1998}. 
Since longer cycles tend to have a larger basin, we find it natural to obtain a smaller value since the second contribution in \cref{eq:barL} is less weighted if the size of basin is not taken into account.

\section{Discussion and conclusions}
\label{sec:discussion}

In this paper, we derived an analytical expression for the average number of limit cycles of length $L$, called $n_L$, 
of a neural network defined by Eq.~\eqref{eq1},
where the connectivity matrix is a random fully connected matrix with asymmetry parameter $\eta$ or $\ee$, thus generalizing the
previous results of Tanaka and Edwards~\cite{Tanaka1980} for the case $L=1$.
We have shown that $n_L \sim (A_L/L) e^{N\Si_L}$ for $N\to\io$, and provided an analytical expression of $\Si_L$. Unfortunately the resulting
expression is difficult to evaluate numerically for generic $L$. We have thus focused
on the case $L \le 3$ and provided results for $\Si_L(\epsilon)$ in that case.  We found that:
\begin{itemize}
\item
$\Sigma_2$ is the largest $\Sigma_L$ for $L=1,2,3$ and for all $\ee$, see figure~\ref{fig:SigmaTheory};
\item
a perturbative expression of $\Si_L$ for $\eta \sim 1-\ee \ll 1$ indicates that $\Si_L \sim e^{-\k L}$ is a decreasing function of $L$, 
Eq.~\eqref{eq:SiLscal},
leading to
the conjecture that $\Sigma_2 > \Sigma_L$, $\forall L$ and $\forall \ee<1$;
\item
all the complexities $\Sigma_L=0$ and
the prefactor $A_L\approx 1$ when $\ee=1$;
\item
for finite $N$, the maximum cycle length is cutoff at $L_c(N,\ee)$, where $L_c(N,\ee=1) \sim e^{\a N}$ with $\a\approx 0.2$.
\end{itemize}
From these results, we conjectured that there exists a critical value $\ee_c\approx 0.797$ (or $\h_c\approx 0.393$)
defined by $\Sigma_2(\ee_c)=\a$, such that:
\begin{itemize}
\item
for $\ee<\ee_c$ or $\h > \h_c$, the average cycle length is dominated by $L=2$, which correponds to the largest complexity $\Sigma_2$;
\item
for $\ee>\ee_c$ or $\h<\h_c$, the largest complexity is still $\Sigma_2$, but the cutoff $L_c(N)$ diverges fast enough that the sum of all cycles
with $2<L<L_c(N)$ is larger than the number of cycles with $L=2$, leading to a divergence of the
average cycle length.
\end{itemize}
Furthermore, we found that for $\ee=1$, the dominant cycles of length $L$ are composed by uncorrelated configurations, which supports the correctness of the 
annealed approximation adopted in \cite{Bastolla1997,Derrida1986}.

Our analysis focused only on the number of cycles, and thus did not take into account the size of the basins of attraction, and for this reason we cannot obtain 
information on the transient time and on the second transition reported in \cite{Bastolla1998}. This is certainly a problem that deserves to be investigated in the future.
Another interesting direction for future work would be to repeat our calculations in the case of finite connectivity random matrices, using the cavity method.

\vspace{1cm} \noindent\textbf{Acknowledgments}
We thank Jacopo Rocchi for interesting discussions.
This project has received funding from the European Research Council (ERC) under the European Union's Horizon 2020 research and innovation programme (grant agreement n. 723955 - GlassUniversality) and (grant agreement n. 694925 - LoTGlasSy).
S. Hwang is supported by a grant from the Simons Foundation (No. 454941, Silvio Franz).

\appendix

\section{Evaluation of Eq.\cref{ComplexityOriginal}}
\label{appendix:EvalActionL}

In this appendix, we present the detailed steps for evaluating $\Sigma_{L}^P$ defined in \cref{ComplexityOriginal}. 
We first compute the two lowest-order terms that together determine both $\Sigma_L^P$ and $A_L^P$ as defined in \cref{eq:partitionFunctionAverage} for a Gaussian distribution.
This result will then be extended in the following subsection to general distribution with non-zero fourth cumulant.

\subsection{Gaussian case}

%
First of all, we present a useful integral representation for the partition function $Z_L$ defined in \cref{eq:partitionFunctionDef}. 
To this end, it is convenient to employ the Fourier representation of a theta function:
\begin{equation}
\th(x) = \int_{-\io}^{\io} \frac{d\l}{2 \pi} \frac{e^{I \l x}}{I\l + \e}
= \int_{-\io}^{\io} \frac{d\l}{2 \pi} \int_{0}^{\infty} d\phi \, e^{I \l ( x -\phi)  }
\ ,
\end{equation}
(where $\e\to 0^+$)
and we define $\DD \l = \frac{d\l}{2\pi (I \l + \e)}$. 
By definition of $\theta$-function, it should be invariant under any scalings of the form $x \to A x$ for positive $A$.
In fact, it can be also checked through any of the above representations for example by taking the joint scaling $\lambda \to A \lambda$, $\phi \to \phi /A$. 
We will shortly exploit this property to reduce the number of free parameters.

Applying this integral representation to each of the theta function appearing in \cref{eq:indicator}, the partition function now takes the form:
\begin{align}
& Z_L^P =
\Measure{P} e^{ \sum_{i,j,t} I\l_i(t) \s_i(t+1) J_{ij} \s_j(t) } \nonumber\\
&= \Measure{P} e^{ \sum_{i > j,t} I S_{ij} \Parenthesis{1 - \frac{\epsilon}{2}} 
	\SquareBracket{\l_i(t) \s_i(t+1) \s_j(t)
		+ i \leftrightarrow j
	} }  e^{
	+ I A_{ij}	\frac{\epsilon}{2}		
	\SquareBracket{\l_i(t) \s_i(t+1) \s_j(t)
		- i \leftrightarrow j
	}
} 
\label{eq:partitionFunctionIntegralRep}
\end{align}
where $i \leftrightarrow j$ denotes the preceding term with $i$ and $j$ exchanged.
The trace $\Measure{P}$ is introduced as a shortcut notation for integrations and summations with respect to variables $\lambda_i$ and $\sigma_i$  with appropriate weights, i.e.,
\begin{align}
\Measure{P}(...) = \sum_{ \ul{\bs} }  \int \prod_{t,i} \frac{d\lambda_i(t)}{2\pi (I \lambda_i(t) + \e)} (...).
\end{align}

As a next step, we proceed to compute the disordered average of $Z_L^P$. 
Specifically, we want to find the  expression for $\Action{L}$ as given by the following equation :
\begin{align}
\QuenchedAverage{Z_L^P} \equiv \Measure{P} e^{N \Action{L}}.
\end{align}
For each independent random variable $X$, which is either $S_{ij}$ or $A_{ij}$, the corresponding term is the average of the form $ \QuenchedAverage{e^{i X m}} $ for a suitable choice of $m$.
This expression is nothing but the characteristic function, which is then $e^{-\frac{J^2 m^2}{2}}$ for the Gaussian distribution.
Using these results for each $S_{ij}$ and $A_{ij}$ yields
\begin{align}
N \Action{L} =& - \sum_{i<j} 
\frac{J^2}2 
\left(1-\frac\ee2\right)^2 
\left(
\sum_{t} \l_i(t) \s_i(t+1) \s_j(t) + i \leftrightarrow j
\right)^2
\nonumber \\
& - \sum_{i<j}   
\frac{J^2}2\left(\frac\ee2\right)^2
\left(
\sum_{t} \l_i(t) \s_i(t+1) \s_j(t) - i \leftrightarrow j
\right)^2.
\end{align}
In order to study the asymptotic behavior of $\PartitionFunc{L}$, it is desirable  to rescale $\Action{L}$ such that it remains to be of $O(1)$.
This can be achieved by using the invariance of $\theta$-function, namely, we employ a transformation $\lambda_i(t) \to \frac{ \sqrt{2 C_\epsilon}}{J \sqrt{N}}  \lambda_i(t)$, where 
$C_\epsilon$ is an arbitrary constant which will be chosen later.
Expanding the squared terms above, we thus have
\begin{align}
 \Action{L} =&  \sum_{i<j} \frac{C_\epsilon}{N^2} 
\Bigg[
\left(1-\frac\ee2\right)^2
\left(
\sum_{t} I\l_i(t) \s_i(t+1) \s_j(t) + i \leftrightarrow j
\right)^2
\nonumber \\	
& + \left(\frac\ee2\right)^2
\left(
\sum_{t} I\l_i(t) \s_i(t+1) \s_j(t) - i \leftrightarrow j
\right)^2
\Bigg]
\end{align}
which is then expanded to
\begin{align}
	\Action{L} =& - \frac{1}{2N^2}  \underset{i\neq j}{\sum} \Bigg\{ 
\sum_{t,s} \l_i(t) \s_i(t+1) \s_j(t)
\l_i(s) \s_i(s+1) \s_j(s)
\nonumber \\ & +\eta
\sum_{t,s}  \l_i(t) \s_i(t+1) \s_j(t)
\l_j(s) \s_j(s+1) \s_i(s)
\Bigg\} \nonumber \\
=& - \frac{1}{2N^2}  \underset{i, j}{\sum} \Bigg\{ 
\sum_{t,s} \l_i(t) \s_i(t+1) \s_j(t)
\l_i(s) \s_i(s+1) \s_j(s)
\nonumber \\ & +\eta
\sum_{t,s}  \l_i(t) \s_i(t+1) \s_j(t)
\l_j(s) \s_j(s+1) \s_i(s)
\Bigg\} \nonumber\\
& + 
\frac{1}{2N^2} (1+\eta){\sum_i} \Bigg\{ 
\sum_{t,s} \l_i(t) \s_i(t+1) \s_i(t)
\l_i(s) \s_i(s+1) \s_i(s)
\Bigg\},	 
\end{align}
where $\eta = (1-\epsilon) /(1-\epsilon + \epsilon^2/2)$ in \cref{eq3} and we have chosen $C_\epsilon = 1/(2 - 2\epsilon + \epsilon^2)$.

Now, the equation can be factored into terms depending only on a single index $i$;
introducing a set of variables 
\begin{align}
R(t,s) &= - \frac1N \sum_i \l_i(t) \s_i(t+1) \l_i(s) \s_i(s+1) \nonumber\\
Q(t,s) &= \frac1N \sum_i \s_i(t) \s_i(s) \nonumber\\
S(t,s) &= \frac1N\sum_i I\l_i(t) \s_i(t+1)  \s_i(s) \nonumber\\
U(t,s) &= \frac{1}{N} \sum_i \l_i(t) \s_i(t+1) \s_i(t) 
\l_i(s) \s_i(s+1) \s_i(s),
\label{OrderParam1}
\end{align}
the terms depending on both indices $i,j$ are completely decoupled: 
\begin{align}
\Action{L} = \ActionCoeff{L}{0}
+ \frac{1}{N} \ActionCoeff{L}{1}  = & \frac{1}{2} \sum_{t, s} \Bigg(
R(t,s)Q(t,s) + \eta S(t,s)S(s,t) \Bigg) + \frac{1 + \eta}{2N} \sum_{t,s} U(t,s).
\label{ZerothOrderFin}
\end{align}

So far, we have rewritten $\Action{L}$ as a function of the newly introduced variables in \cref{OrderParam1}.  
These relations can be implicitly imposed by employing a set of delta functions.
For example, for each variable $Q(t, s)$
we introduce a trivial identity as a double integral
\begin{align}
& 1 = \frac{N}{2\pi}\int_{Q(t,s) \hat{Q}(t,s)} d Q(t,s) d \hat Q(t,s) e^{ I N \hat{Q}(t,s) ( Q(t,s)} e^{- \sum_i \s_i(t) \s_i(s) )}.
\end{align}
Before writing a complete expression that will be unmanageably large, let us make a couple of simplifications.

After the substitutions, the asymptotic behavior of the remaining integrals are evaluated via the saddle point method.
Since $U(t,s)$ appears in the next-to-leading order $O(N^{-1})$, the exponential rate $\Sigma_L^P$  cannot be perturbed by the presence of this term.
Thus, we must have 
\begin{align}
\hat U(t,s)^* &=0.
\end{align}

Additionally, we can reduce the number of delta functions by observing that $\Action{L}$ in \cref{ZerothOrderFin} is mostly linear in certain variables. 
Removing redundant variables through relations such as $I R(t,s) =\hat{Q}(t,s)$ or $Q(t,s) = Q(s,t)$, 
we find 
\begin{align}
\PartitionFunc{L} =& \left(
\frac{N i }{2\pi}
\right)^{\frac{L(L-1)}{2} }
\left(
\frac{N i }{2\pi}
\right)^{\frac{L^2}{2} }
\int_{R, Q, S} 
e^{ N \Sigma_L^P} e^{\ActionCoeff{L}{1}},
\label{PartitionFunctionSaddlePoint}
\end{align} 
where $\Sigma_L^P$ is given in \cref{ComplexityOriginal}.
The multiplicative prefactor $e^{\ActionCoeff{L}{1}}$ is then evaluated by computing
\begin{align}
U(t,s) = \Avr{
	\l(t) \s(t+1) \s(t) 
	\l(s) \s(s+1) \s(s)
}_P
\label{EquationForU}
\end{align}
where
\begin{align}
\Avr{\cdots}_P =& \frac{1}{\mathcal{Z}_L^P} \MeasureSite{P}
e^{
	-\frac{1}{2} \sum_t \lambda(t)^2} e^{ 
	- \sum_{t>s} Q(t,s) \l(t) \s(t+1) \l(s) \s(s+1)
} \nonumber \\ & e^{ \sum_{t>s} R(t,s) \s(t) \s(s) +  \sum_{t,s} \eta S(t,s)  I\l(t) \s(t+1) \s(s) 
} \Parenthesis{\cdots}.
\label{AngleBracket}
\end{align}

Finally, after determining the saddle points of $Q(t,s)$, $R(t,s)$ for $t>s$ and $S(t,s)$, we arrive at 
\begin{align}
\PartitionFunc{L} =& 
\frac{e^{ N \Sigma_L^P} e^{\ActionCoeff{L}{1}}}{\sqrt{ | \det H(\{Q(t,s), R(t,s) ,S(t,s)			
		\})|}},
\label{PartitionFunctionFinal}
\end{align}
where 
$ H(\{Q(t,s), R(t,s) ,S(t,s)			
\} $ is the Hessian matrix constructed at the saddle point.


\subsection{General case}
As a next step, let us consider a general case for symmetric distributions with non-zero fourth-order cumulant.
Previously, we have pointed out that each disorder average is of the form $ \QuenchedAverage{e^{i X m}} $. 
As we consider a general case, this term allows a cumulant expansion of the form 
\begin{align}
\log  \QuenchedAverage{e^{i X m}} = - J^2 \frac{m^2}{2} + \frac{\kappa}{4} m^4 + O(m^6),
\end{align} 
where we did not take a conventional denominator $4!$ but rather use $4$ for simplicity.
As previously argued in the case of a Gaussian distribution, the same scaling $\lambda_i(t) \to \frac{ \sqrt{2 C_\epsilon}}{J \sqrt{N}}  \lambda_i(t)$ is needed to make $ \Action{L} $ of $O(1)$. 
Thus, the higher order contributions due to the presence of $\kappa$ appears only as corrections of order $O(N^{-1})$.

Now, let us evaluate the additional term explicitly: 
\begin{align}
\mathcal{G} =& \sum_{i<j} 
\frac{ C_\epsilon^2 \kappa }{J^4 N^2 } 
\Bigg[
\left(1-\frac\ee2\right)^4 
\left(
\sum_{t} I\l_i(t) \s_i(t+1) \s_j(t) + i \leftrightarrow j
\right)^4 \nonumber \\ &
+ \left(\frac\ee2\right)^4
\left(
\sum_{t} I\l_i(t) \s_i(t+1) \s_j(t) - i \leftrightarrow j
\right)^4
\Bigg].
\end{align}

Similarly to the Gaussian case, $\mathcal{G}$ can be written in a succinct way by introducing a set of variables 
\begin{align}
V_x(t_1, t_2, t_3, t_4) &= \frac1N \sum_i 
\Parenthesis{
	\prod_{g=1}^{x} \l_i(t_g) \s_i(t_g+1)
}
\Parenthesis{
	\prod_{g=x+1}^{4} \s_i(t_g)
}
\label{OrderParam2}	
\end{align}
for $x= 0, 1, \cdots, 4$.
Thus, $\mathcal{G}$ reads
\begin{align}
N \mathcal{G} =  \frac{ C_\epsilon^2 \kappa }{J^4 }  \SquareBracket{
	X_1 + X_2 + X_3
} + O(N^{-2}),
\label{FirstCorrection}
\end{align}
where 
\begin{align}
X_1 = & \Parenthesis{\left(1-\frac{\epsilon }{2}\right)^4+\frac{\epsilon ^4}{16}}
\sum_{t_1, t_2, t_3,t_4} V_4(t_1,t_2,t_3,t_4) V_0(t_1,t_2,t_3,t_4),   
\end{align}
\begin{align}
X_2 = & 4 \Parenthesis{\left(1-\frac{\epsilon }{2}\right)^4-\frac{\epsilon ^4}{16}} 
\sum_{t_1, t_2, t_3,t_4} V_3(t_1,t_2,t_3,t_4) V_1(t_4,t_1,t_2,t_3),   
\end{align}
and 
\begin{align}
X_3 = & 3 \Parenthesis{\left(1-\frac{\epsilon }{2}\right)^4+\frac{\epsilon ^4}{16}
} \sum_{t_1, t_2, t_3,t_4} V_2(t_1,t_2,t_3,t_4) V_2(t_3,t_4,t_1,t_2).
\end{align}
Thus, once the saddle point of $\Sigma_L^P$ in \cref{ComplexityOriginal} is determined, 
we now include the contribution of $\mathcal{G}$ to the partition function \cref{PartitionFunctionFinal}:
\begin{align}
\PartitionFunc{L} =& 
\frac{e^{ N \Sigma_L} e^{\mathcal{H}} }{\sqrt{ | \det H(\{Q(t,s), R(t,s) ,S(t,s)			
		\})|}},
\label{PartitionFunctionFinal2}
\end{align}
where $ \mathcal{H} = \ActionCoeff{L}{1} + \mathcal{G}$ and $V_x(t_1,t_2,t_3,t_4)$'s in $\mathcal{G}$ are determined via
\begin{align}
V_x(t_1,t_2, t_3, t_4)  = \Avr{
	\Parenthesis{
		\prod_{g=1}^{x} \l(t_g) \s(t_g+1)
	}
	\Parenthesis{
		\prod_{g=x+1}^{4} \s(t_g)
	}
}_P.
\label{EquationForV}
\end{align}

\section{Perturbative approach for $\PartitionFunc{L}$ around $\epsilon=1$}
\label{appendix:Perturbation}
To understand $\Sigma_L^P$ close to $\epsilon =1$, we applied perturbation theory, in which the variables appear as a power series of $\eta$.
To this end, we need to evaluate a huge number of integrals as a result of the higher-order expansions in  \cref{ComplexityOriginal}.
Nevertheless, this can be carried out systematically exploiting the observation that every term appearing in the expansion of $\log \mathcal{Z}_L^P$ should always be of the following factorized form
\begin{align} 
\Avr{\prod_t \sigma(t)^{e(t)} \lambda(t)^{f(t)}}_P  \equiv & \,  \MeasureSite{P}
e^{
	-\frac{1}{2} \sum_t \lambda(t)^2 
} \prod_t \sigma(t)^{n_1(t)} \lambda(t)^{n_2(t)},
\label{PerturbationIntegral}
\end{align}
for some positive integers $n_1(t)$ and $n_2(t)$.
By solving the integrals one by one for each $t$, it is then easy to verify that this integral is nonzero only if all the $n_1(t)$'s are even and $n_2(t)$'s are either odd or zero.
The second condition is established by the following identity:
\begin{align}
G_k \equiv \int_{-\io}^{\io} \frac{d\l}{2 \pi} \int_{0}^{\infty} d\phi \, e^{- I\phi \l  } e^{-\frac{1}{2} \lambda^2} \lambda^k  = \frac{i 2^{\frac{k}{2}-2} \left((-1)^k-1\right) \Gamma
	\left(\frac{k}{2}\right)}{\pi }
\label{LambdaIntegral}
\end{align}
for arbitrary positive integers $k$.

For the symmetric boundary condition $P=1$, by the definition of cycles, translation invariance holds. 
As a result, the number of independent variables can be dramatically decreased.
Here, we assume that the saddle point in the vicinity of $\eta =0$ allows the following expansion:
\begin{align}
X(t) = X_0(t) + \eta X_1(t) + \eta^2 X_2(t) + \cdots,
\end{align}
where $X$ is any of $\{Q, R, S\}$. 

Now, we briefly sketch how to determine the first-order correction.
By expanding \cref{ComplexityOriginal} up to $ O(\eta) $, we are led to compute the following averages:
\begin{align}
&	\eta \Bigg\langle \sum_{t>s} Q_1(|t-s|) I\l(t) \s(t+1) I\l(s) \s(s+1) 
+ \sum_{t>s} R_1(|t-s|) \s(t) \s(s) + \\ & \sum_{t,s} S_0(t-s)  I\l(t) \s(t+1) \s(s) \Bigg\rangle_+  + o(\eta) 
\end{align} 
Using the criterion specified below \cref{PerturbationIntegral}, it is easy to show that the first two terms vanish for every pair of $t>s$.
Similarly, one can see that the third term does not vanish only when $t+1 = s$.
Thus, collecting all the nonzero contributions in \cref{ComplexityOriginal}, we have the following 
\begin{align}
(\Si_L^+)_1  &=
\frac12 
\sum_{t,s} S(s-t) S(t-s) + \sqrt{\frac{2}{\pi}} L \, S(-1) + O(\eta).
\end{align}
Thus, the corresponding saddle point from the above action is readily found as
$S_0(t) = \sqrt{\frac{2}{\pi}} \delta_1$, which then implies 
$ \Sigma_1^+ = \frac{1}{\pi} \eta + O(\eta^2) $, $ \Sigma_2^+ = \frac{2}{\pi} \eta + O(\eta^2) $ and $ \Sigma_L^+  =O(\eta^2) $ for $L > 2$.

Repeating the procedure up to fourth-order coefficients in $\eta$, the series is found to be
\begin{align}
Q(t) &= \frac{2}{\pi -2 } (\delta_{t,2}( 1 + \delta_{L,4})   ) \eta+ \frac{8 (\pi -1)}{(\pi -2)^2 \pi } (\delta_{t, L-4}(1  + \delta_{L, 8} )) \eta^2 + O(\eta^3)\nonumber \\
R(t) &= O(\eta^3) \nonumber \\ 
S(t) &= \sqrt{\frac{2}{\pi}} \delta_{t,1} + \frac{2\sqrt{2} }{\pi^{3/2}}  \delta_{t,3} \eta + O(\eta^2).
\end{align}

Due to the lack of translation symmetry for $P=-1$, the perturbative analysis is more involved. 
Instead of considering one-time variable, we need to keep two time quantities in \cref{ComplexityOriginal}.
Apart from that, we can proceed similarly to the case of $P=1$. 
Within our computational capacity, we managed to perform 2nd order perturbation theory.

\section{Computing $\PartitionFunc{L}$ at $\epsilon=1$}
\label{appendix:EvalPrefactor}
At $\epsilon =1$, the saddle point analysis for \cref{ComplexityOriginal} becomes relatively straightforward. 
Since $S(t,s)$ does not appear in the action, only the other two matrices $Q(t,s)$ and $ R(t,s)$ should be extremized.
Namely, the partition function \cref{PartitionFunctionSaddlePoint} for $\eta=0$ reads
\begin{align}
\PartitionFunc{L} =& \left(
\frac{N i }{2\pi}
\right)^{\frac{L(L-1)}{2} }
\int_{R, Q} 
e^{ N \Sigma_L^P} e^{\ActionCoeff{L}{1} + \mathcal{G}},
\label{PartitionFunctionAtOne}
\end{align} 
where 
\begin{align}
\Si_L^P &=
- \sum_{t > s} 
R(t,s) Q(t,s)
+ \log
\left( \mathcal{Z}_L^P
\right),
\end{align}
and
\begin{align}
\mathcal{Z}_L^P & = 
\MeasureSite{P}
e^{
	-\frac{1}{2} \sum_t \lambda(t)^2 } e^{ 
	- \sum_{t>s} Q(t,s) \l(t) \s(t+1) \l(s) \s(s+1)} e^{
	+ \sum_{t>s} R(t,s) \s(t) \s(s) 
}.
\end{align}
From this, it is straightforward to see that there exists a saddle point corresponding to $R(t,s) = 0$ and $Q(t,s) =0$ for all $t>s$.

At this saddle point, the integrals for $\Avr{\cdots}_P$ defined in \cref{AngleBracket} is simplified to
\begin{align}
\Avr{\cdots}_P = \MeasureSite{P}
e^{
	-\frac{1}{2} \sum_t \lambda(t)^2 
} \Parenthesis{\cdots},
\end{align}
where we have used the relation $\mathcal{Z}_L^P = 1$ at $Q(t,s)=R(t,s) = 0$.

First, let us determine $U(t,s)$, $V_x(t_1, t_2, t_3, t_4)$ using \cref{EquationForU,EquationForV}.
For $U(t,s)$, we need to evaluate the following integral:
\begin{align}
U(t,s) = U(s,t) = \Avr{
	\l(t) \s(t+1) \s(t) 
	\l(s) \s(s+1) \s(s)
}
\end{align}
In order to have a nonzero contribution, the spin variables should be of even power, while the lambda variables of odd power. 
This can be achieved only when $L=2$ and $t = s+1$.
Using the identity \cref{LambdaIntegral}, $U(s+1,s)$ is evaluated to $- \frac{ 2 P}{\pi} \delta_{s, s+2}$.
Similarly, one can show that 
\begin{align}
X_1 = 0, \quad X_2 =0
\end{align}
and
\begin{align}
X_3 =\frac{6}{\pi^2} \Parenthesis{
	\delta_{L, 2}  + 2P \delta_{L, 4}
}.
\end{align}
Plugging these solutions into \cref{FirstCorrection}, \cref{ExponentialCorrectionAtOne} is derived.
Surprisingly, we find that $\ActionCoeff{L}{1}$ and $\mathcal{G}$ are mostly zero except for special cases $L=2$ and $L=4$.

Next, let us compute the determinant of Hessian matrix at the saddle point. 
For later convenience, we introduce $\hat Q(t,s) = i R(t,s) $.
Expanding $\log \mathcal{Z}_L^P$ up to second order, we find that 
this step is equivalent to performing the following Gaussian integral:
\begin{align}
M_P(L) &= \int \prod_{t>s} \left[
\frac{dQ(t,s) 
	d \hat{Q}(t,s)}{2\pi} 
\right] 
e^{
\sum_{t>s} (
- \frac{\hat{Q}(t,s)^2}{2} 
+ i \hat{Q}(t,s) Q(t,s)
- \frac{2}{\pi} i \hat{Q}(t,s) Q(t-1, s-1) P^{\delta_{t,0}}  
)} \nonumber \\
&= \int \prod_{t>s} \left[
\frac{ dQ(t,s) }{\sqrt{2\pi}} 
\right] 
\exp \left[
\sum_{t>s} 
- \frac{ \left(Q(t,s) - \frac{2}{\pi} Q(t-1,s-1) P^{\delta_{t,0}} 	\right)^2 }{2} 
\right].
\end{align}
The last equation implies that the spin configurations only interact with others having the same two-time distance $|t-s|$.
Thus, for each distance $d = |t-s|$, this integrand is simply a Gaussian integral on a linear chain.
Specifically, they form $\frac{L-1}{2}$ $L$-chains if $L$ is odd, while $\frac{L-2}{2}$ $L$-chains and one $L/2$-chain.

For $P=1$, the couplings between two adjacent nodes are uniform, and thus the corresponding quadratic form is circulant. 
Using the well-known results in the theory of circulant matrices, we find that the integral on the linear chain with length $L$ is evaluated to 
\begin{align}
C_L = \frac{\pi^L}{\pi^L - 2^L},
\end{align}
which results in
\begin{align}
M_+(L) = \left\{\begin{array}{lr}
(C_L)^{\frac{L-1}{2}} , & \text{for  odd } L\\
(C_L)^{\frac{L-2}{2}} C_{L/2}, & \text{for even } L
\end{array}\right.	
\end{align}

For $P= -1$, it turns out that the integral for each $L$ chain gives the same result as the one for $P=1$. 
The only difference comes from the chain with length $L/2$ if $L$ is even, in which case, there is one positive coupling constant instead of a negative one. 
For this case, one can compute the integral of the form
\begin{align}
D_L = \frac{\pi^L}{\pi^L + 2^L},
\end{align}
and thus
\begin{align}
M_-(L) = \left\{\begin{array}{lr}
(C_L)^{\frac{L-1}{2}} , & \text{for  odd } L\\
(C_L)^{\frac{L-2}{2}} D_{L/2}, & \text{for even } L
\end{array}\right..
\end{align}
Finally, inserting both contributions into \cref{PartitionFunctionFinal} completes our analysis:
\begin{align}
\overline{Z_L}= M_P(L) \exp(\mathcal{H} ),
\end{align}
where $ \mathcal{H} = \ActionCoeff{L}{1} + \mathcal{G}$.

\begin{figure}
	\begin{center}
	\includegraphics[width=.75\textwidth]{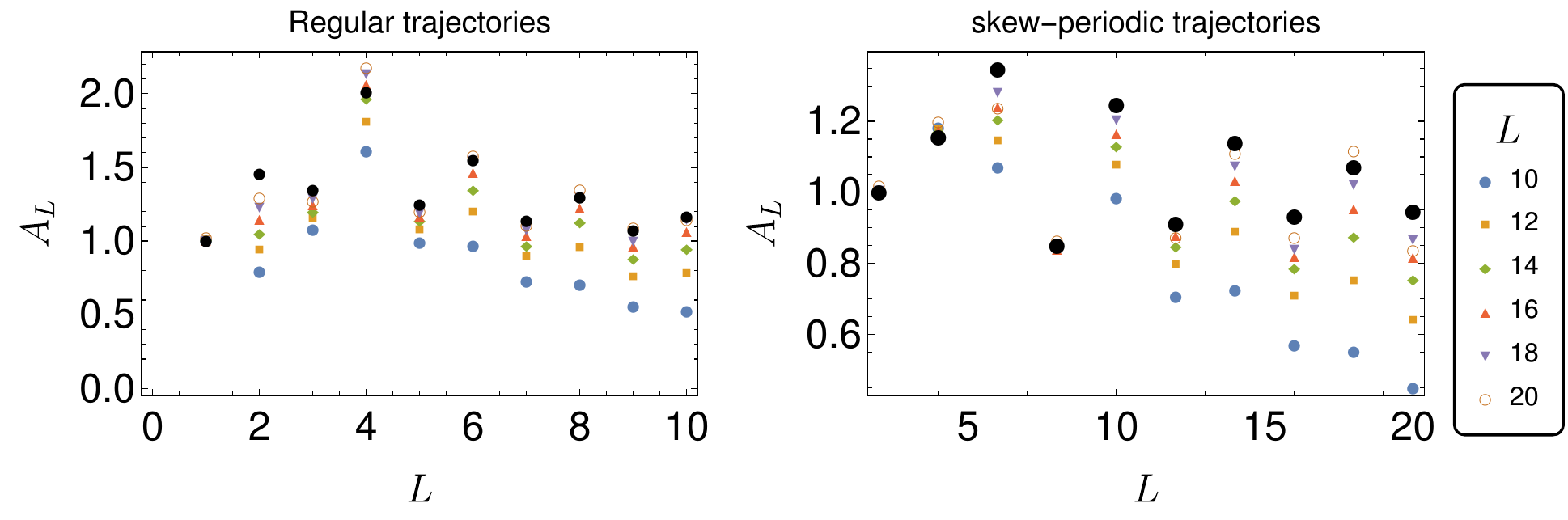}
	\end{center}
	\caption{Simulation results for $A_L^P$ for various $L = 10, 12, 14, \cdots , 20$. A Gaussian distribution is used to generate coupling matrices. The data points (small symbols) seem to approach the theoretical value, which is marked by the black large dots with notable exceptions for $L=2$ or $L=4$.
	}
\label{fig:SimulationAL}
\end{figure}

In order to corroborate our results, we have performed an extensive simulation to determine $A_L^P$ for system sizes up to $20$.
In \cref{fig:SimulationAL}, we show the values of $A_L^P$ for different sizes of $L$.
This shows that simulation data are mildly scattered around the theoretical values. 
In some cases, we also observe a clear trend with the data converging to the theoretical point.
However, this is not always true. 
Especially for $P=1$ at $L=4$, the simulation results clearly overshoot the theoretically predicted point (black dot).
We attribute this deviation to a finite size effect.
In fact, we will provide a strong evidence to support this claim in \cref{appendix:DirectInt} by directly computing $A_2^P$ through an exact integration.

\section{Direct integration of $Z_2^P$ at $\epsilon=1$}
\label{appendix:DirectInt}

In \cref{appendix:EvalPrefactor}, we have observed some disagreements between simulation data and the theoretical predictions for $A_L^+$ especially for $L=2$ and $L=4$. 
Here, our direct integration will show that this discrepancy is simply attributed to a finite size effect in the case of $L=2$. 
For $ L=4 $, it is difficult to repeat the same procedure. 
Nevertheless, we believe the same scenario should apply as well.

The choice of Gaussian distribution makes the analysis easier. 
In this case, \cref{PartitionFunctionAtOne} is exact since no truncation in the cumulant expansion has been made. Recovering the exact integral domains, the partition function reads
\begin{align}
\PartitionFunc{2} =& A_2^P = \left(
\frac{N  }{2\pi}
\right)
\int_{-i\infty} ^{i \infty } \int_{-\frac{N-2}{N}}^{\frac{N-2}{N}} dR dQ \exp \SquareBracket{
	-NQ R + N \log \mathcal{Z}_2^P
}	,
\end{align}
where 
\begin{align}
\mathcal{Z}_2^P = \sum_{\sigma = \pm 1} \Phi_2 \Parenthesis{
	\frac{P N}{N-1} \Parenthesis{
		Q \sigma - \frac{1}{N} 
	}
} e^{R \sigma}
\end{align} 
with 
$\Phi_2(x) = \frac{\pi + 2 \sin ^{-1}(x) }{2 \pi }$.
This function $ \Phi_2(x) $ corresponds to two times the probability that two correlated Gaussian random variables with correlation parameter $x$ are both positive. 
Note that the integral domain of $Q$ should be now considered as a discrete sum rather than an integral.
Also, we have excluded the trivial case by only considering $\Parenthesis{
	-\frac{N-2}{N}, \frac{N-2}{N}
}$. 
Finally, expanding the term 
$ (\mathcal{Z}_2^P )^N$ using the binomial theorem and performing the delta-function integrals with respect to $R$, we arrive at 
\begin{align}
\PartitionFunc{2} = \sum_{k=1}^{N-1} \binom{N}{k} \Phi_2\Parenthesis{
	-\frac{P (-2 k  +N  +1)}{N-1}
}^{N-k} 
\Phi_2\Parenthesis{
	-\frac{P (2 k -N  +1)}{N-1}	
}^k.
\label{ParitionExactGaussianL2}
\end{align}

In order to study the same problem for arbitrary distributions, we need to come up with a different approach.
This relies on the fact that our formalism is based on truncation of cumulant expansion.
Surprisingly, we can construct a powerful formalism that works for arbitrary distributions for the case of $\epsilon=1$.

Now, let us compute the number of cycles of period 2, i.e., $ \PartitionFunc{2} $, using a new approach.
By symmetry, we may focus on the configuration with $\sigma_i = 1$ for all $i = 1, ..., N$.
At the next time step, we can imagine that the configuration will evolve to another with $k$ positive $\sigma_i$'s and $(N-k)$ negative $\sigma_i$'s for some $k$.
Certainly, there are $\binom{N}{k} $ different configurations corresponding to such event.
As each choice gives an identical contribution, 
let us reorder the spin indices such that first $k$ spins are positive while $N-k$ spins are negative.
The case $k=0$ and $k=N$ are trivial and they appear with weight 1.
The probability will depend if the spin at the first step is positive or negative.
Let us call these probabilities
$U_\pm^P(k)$.
With these definitions, the mean number of cycles should satisfy the following formula
\begin{equation}
\overline{Z_2^P} = \sum_k  \binom{N}{k} \Parenthesis{U_+^P (k)}^k \Parenthesis{U_-^P (k)}^{N-k}
\label{eq:ExactAnyDist}
\end{equation}

Now, let us determine $U_+^P(k)$ for neurons $1 \le  i \le k$.
For each neuron $i$, there is an associated quenched synaptic coupling vector $J_{ij}$ with $J_{ii} = 0$.
According to our conditioning, 
the coupling vector should satisfy
\begin{align}
\Parenthesis{\sum_{j = 1}^{k} J_{ij} + \sum_{j = N}^{k+1} J_{ij}} > 0,
\end{align}
for $1 \le  i \le k$.
Moreover, the condition for the path being closed can be written as
\begin{align}
\Parenthesis{\sum_{j = 1}^{k} J_{ij} - \sum_{j = N}^{k+1} J_{ij}} P > 0.
\end{align}
Luckily, both conditions can be written only in terms of two quantities, i.e., $ a \equiv \sum_{j = 1}^{k} J_{ij}  $ and $ b \equiv  \sum_{j = N}^{k+1} J_{ij}$ and their probability densities are given by the $(k-1)$-fold (due to the condition $J_{ii} =0$) and $(N-k)$-fold convolution of the coupling distribution $\mathcal{P}(J)$
~\footnote{Note that this distribution is not identical to $P(x)$ which is a PDF of $S_{ij}$ and $A_{ij}$.}.
Putting them together, one can find
\begin{align}
U_+^P(k) &= \frac{\int d a d b \, \mathcal{P}_{k-1}(a) \mathcal{P}_{N-k}(b)   \theta( P(a-b)) \theta(a+b)}{\int d a d b \, \mathcal{P}_{k-1}(a) \mathcal{P}_{N-k}(b)   \theta(a+b)}  \nonumber \\
&= 2 \int d a d b \, \mathcal{P}_{k-1}(a) \mathcal{P}_{N-k}(b)   \theta( P(a-b)) \theta(a+b),
\label{eqApp1}
\end{align}
where $\mathcal{P}_x(a)$ refers to the $x$-fold convolution of $\mathcal{P}(J)$ and the denominator of the first equation is introduced due to the condition on being a positive bit.
Along the same line, one can easily find
\begin{equation}
U_-^P(k)= 2 \int d a d b \, \mathcal{P}_{k}(a) \mathcal{P}_{N-k-1}(b)   \theta( P (a-b) ) \theta(-a-b).
\label{eqApp2}
\end{equation}
Since $U_+^P(k)$ and $U_-^P(k)$ can be determined for any distributions, the exact solution \cref{eq:ExactAnyDist} can be obtained.

Now, let us consider two interesting special cases, i.e., i) a Gaussian distribution and ii) a binary distribution.
For the Gaussian distribution, one can easily check that Eqs. (\ref{eqApp1}) and (\ref{eqApp2}) correspond to two times the probability that both elements of a random vector drawn from a bivariate Gaussian with a certain correlation $\rho_\pm$ are positive, where the correlations are given by $\rho_{+}
= P \frac{2 k-N-1}{N-1}$ and $ \rho_{-}
= P \frac{-2 k+N-1}{N-1} $, respectively.
Note that this reproduces exactly the formula we obtained using a different approach \cref{ParitionExactGaussianL2}.

Let us focus our attention to the binary distribution.
In this case, $\mathcal{P}_x(a)$ corresponds to a binomial distribution.
Expanding $\mathcal{P}_x(a)$ as a binomial summation, we arrive at
\begin{equation}
U_+^P(k)= \frac{2}{2^{N-1}} \sum_{a=0}^{k-1} \sum_{b=0}^{N-k} \binom{k-1}{a} \binom{N-k}{b} \theta(2(a-b) + N -2k +1) \theta(2(a+b) -(N-1))
\label{UBinaryPlus}
\end{equation}
and
\begin{equation}
U_-^P(k)= \frac{2}{2^{N-1}} \sum_{a=0}^{k} \sum_{b=0}^{N-k-1} \binom{k}{a} \binom{N-k-1}{b} \theta(2(a-b) + N -2k -1) \theta(-2(a+b) +(N-1)).
\label{UBinaryMinus}
\end{equation}

\begin{figure*}
	\begin{center}
	\includegraphics[width=.75\textwidth]{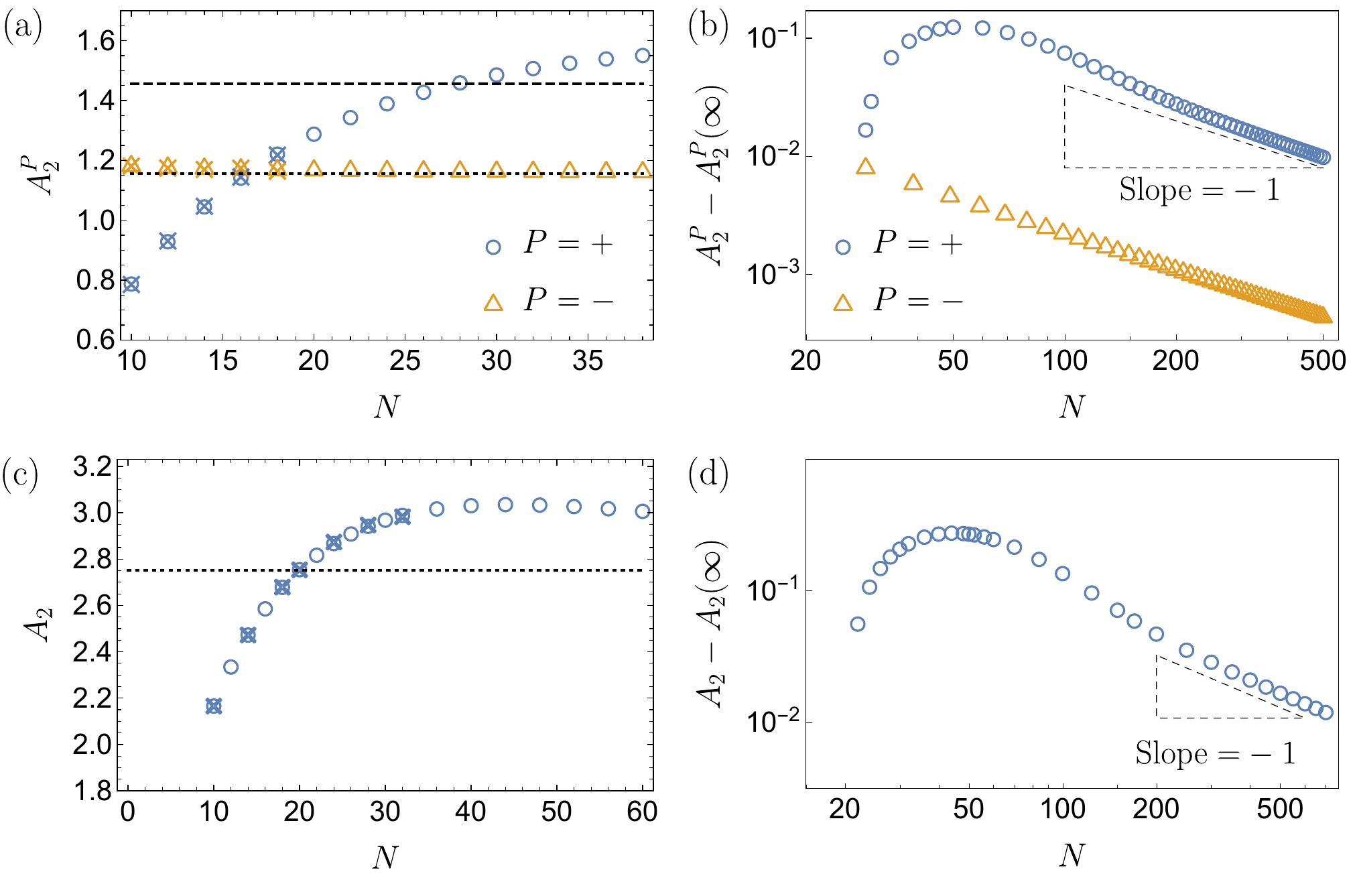}
	\end{center}
	\caption{
		(a) Estimations of $A_2^P$ from the direct integration in \cref{eq:ExactAnyDist} (empty symbols) and the numerical simulations for $N < 20$ (crosses).
		The black dashed line and the dotted line indicate the asymptotic values $\frac{\pi }{\pi-2} e^{-2/\pi}$ and $\frac{\pi }{\pi+2} e^{2/\pi}$, respectively. For $P = +1$, the data points clearly overshoot the theoretical prediction.
		(b) Double-logarithmic plot of $A_2^P - A_2^P(\infty)$ as a function of $N$. 
		As predicted by our formalism, both curves decays as $N^{-1}$, corroborating the validity of our theoretical frameworks.
		(c) and (d) Same procedure is repeated for the binary distribution using \cref{eq:ExactAnyDist,UBinaryPlus,UBinaryMinus}.
	}
	\label{fig:DirectIntegral}
\end{figure*}

In \cref{fig:DirectIntegral} (a), we plotted the exact solutions (open symbols) as predicted by \cref{eq:ExactAnyDist} as well as the results of numerical simulations (crosses) for the case of Gaussian couplings.
Since the numerical errors are negligible compared to the size of the symbols, the error bars are omitted.
What is surprising especially for $P =1$ is that the data points, as a function of $N$, overshoots the asymptotic result $A_2^+(N = \infty) = \frac{\pi}{\pi-2} e^{-2/\pi}$ around $N \simeq 25$.
However, it turns out that it is simply because of a strong finite size correction. 
To illustrate this point, we have drawn the difference between $ A_2^+ $  and its asymptotic value $ A_2^+(N = \infty) $ (See \cref{fig:DirectIntegral} (b)). 
The figure shows that both errors for $P =\pm$ decay as $N^{-1}$ as predicted by our formalism.
In \cref{fig:DirectIntegral} (c) and (d), we repeat the same analysis for the binary distribution, in which we found the same pattern.

To describe the strong finite size correction, one may extract the $1/L$-correction from the exact solution \cref{eq:ExactAnyDist} of the form $A_2 = A_2^{(0)} + A_2^{(1)}/N + O(1/N^2)$.
We found that $ A_2^{(1)} / A_2^{(0)} =  \frac{4 (4+(\pi -2) (\pi -1) \pi )}{(\pi -2)^3 \pi ^2} \simeq 3.18$. Thus, to obtain a reliable estimate of the asymptotic value of $A_2$ within few percents of error, one needs to increase the system size to $N \sim O(10^2)$. 
In \cref{fig:DirectIntegral}, one can indeed see the reasonable convergence in the range of $N \sim O(10^2)$.

\section{More on the average cycle length}
\label{appendix:MoreOnCycleLength}

\begin{figure}
	\begin{center}
		\includegraphics[width=.85\textwidth]{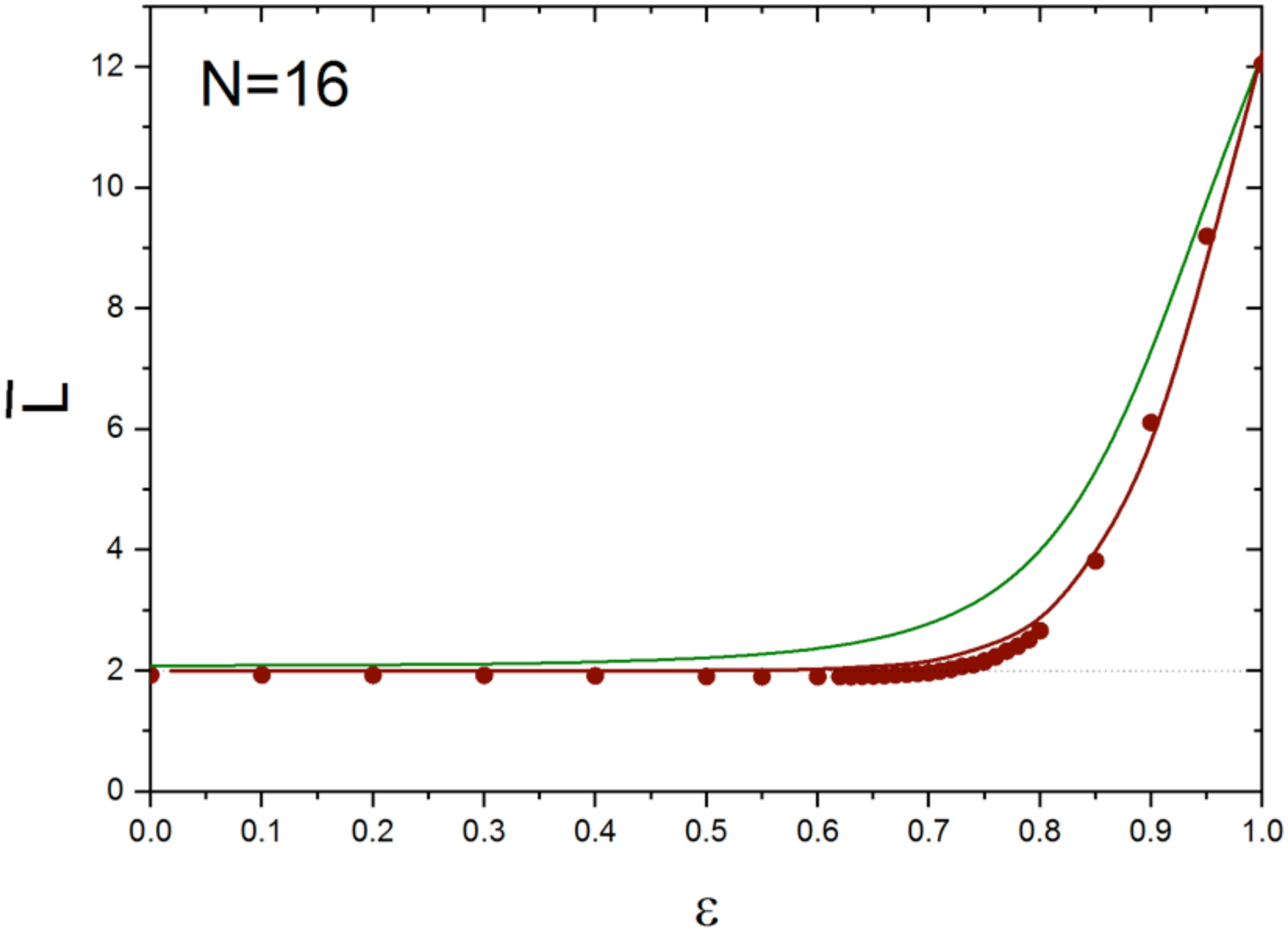}
	\end{center}
	\caption{
		Mean cycles length, $\bar L$, as a function of $\epsilon$ for the case $N=16$. Red dots: numerical simulations. Green line: Equation 
		(\ref{eq:barL2}) with $\alpha$ kept fixed at its $\epsilon$=1 value,  $\alpha$=0.21. Red line: Equation (\ref{barL}) with $\alpha(\epsilon)$ 		determined from the exponential $N$ dependence of the cut-off $L_c(\epsilon, N)$ (see Figure \ref{f8}).}
	\label{f9}
\end{figure}

The derived value for $\epsilon_c$ strongly depends on the assumption that $\alpha$ does not depend on $\epsilon$. 
To check this hypothesis, we need a robust derivation of these quantities, which relies on a good determination of $L_c$. 


Given the explicit expression for $n_L$ (see Eq.~\eqref{func}) and the $N$ dependence of $L_c(N) = e^{\alpha N +\beta}$, with $\alpha$ and $\beta$ obtained by fits such as those ones exemplified in the insets of the right panel of figures \ref{simulazioni4} and \ref{simulazioni5}, the quantity $\bar L$ (Eq.~\eqref{eq:barL}) can be estimated by approximating the sum with integrals and keeping the leading and sub-leading terms in $L_C$ as follows:  

\begin{equation} 
\label{barL}
\fbox{ $ \displaystyle
\bar L = \frac{ 2 + \large [ \Gamma(\frac{3}{2})L_c - 3 \large ] e^{-\Sigma_2 N} }{1+\large [\frac{3}{2} \log(\frac{L_c}{3}) - \gamma \large ] e^{-\Sigma_2 N}} .
$}
\end{equation}
At $\epsilon$=1, where $\Sigma_2$=0, $\alpha$=0.21 and $\beta$=0.93, this equation for the case $N$=16 gives $\bar L$=12.2, which compares favourably with the simulation value 12.1. 

If we now keep the parameters $\alpha$ and $\beta$ fixed at all $\epsilon$, we can compute the $\epsilon$ dependence of $\bar L$. As an example, this quantity is reported for $N$=16 as a green line in Figure \ref{f9}, and it clearly fails to satisfactorily represent the whole $\epsilon$ dependence of $\bar L$ as obtained by the numerical simulation (full red dots). This is a clear indication of the failure of the hypothesis that $L_c$ does not depend on $\epsilon$.

\begin{figure}
	\begin{center}
		\includegraphics[width=.85\textwidth]{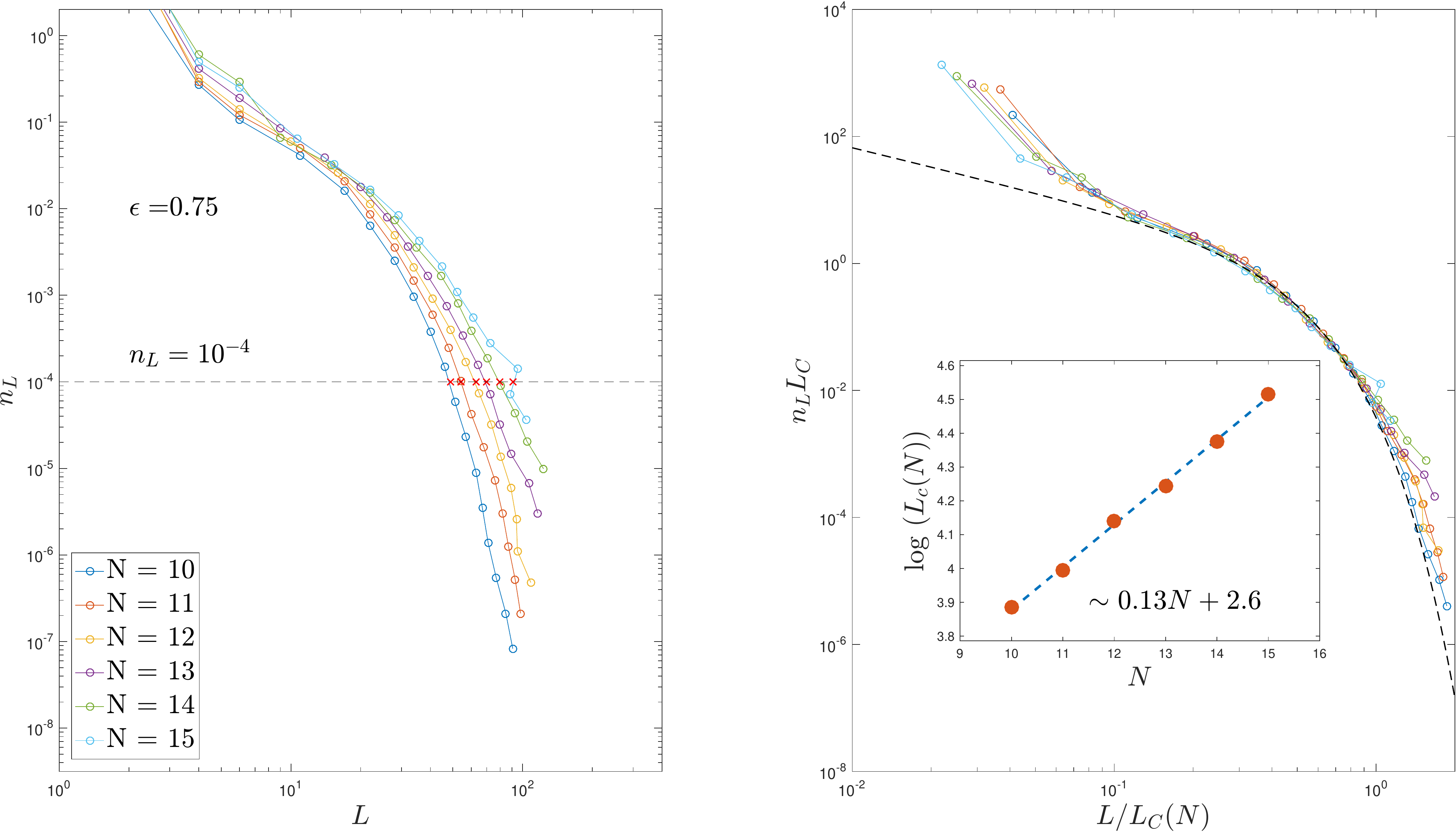}
	\end{center}
	\caption{Left panel: The number of cycles of length $L$, $n_L$, is reported as a function of $L$ for different $N$ values and, as an example, 		     for the case $\epsilon=0.75$. The large $L$ portion of the data recall the behavior found for $\epsilon=1$, with a clear cutoff, while 		     at small $L$ a second $L$-scale appears. The value of the cut-off, $L_c(\epsilon,N)$, is determined by the condition $n_L=10^{-4}$.
	             Right panel: the scaled quantity $n_L L_c$ is reported a s a function of $L_c/L$. A good collapse is observed at large $L$ values. 		    The inset shows the $N$ dependence of $L_c$, demonstrating the validity of the relation $L_c \approx exp(\alpha(\epsilon) N +			    \beta(\epsilon))$. $\alpha(N)$ turns out to decrease on decreasing $\epsilon$.}
	\label{f8}
\end{figure}

To determine the $\epsilon$ dependence of $\alpha$ and $\beta$, we simulated $n_L(\epsilon,N)$, reported at $\epsilon$=0.75 in Figure \ref{f8} as an example. We notice that {\it i)}  for $\epsilon < 1$ the function $x e^{-x^{3/2}}$ no longer describes the data; {\it ii)} beside $L_c$, a second, shorter ``scale'' appears in the description of $n_L$; {\it iii)} a cut off $L_c$ can still be introduced. As we cannot rely on the fit for the determination of $L_c$, we determine this quantity from the condition $n_L$=10$^{-4}$. This yields $L_c$ values up to an unknown proportionality factor, which is fixed at all $\epsilon$ by using the known value at $\epsilon$=1. As a consistency check, the obtained values for $L_c(\epsilon, N)$ are entered in Eq.~\eqref{func} and the result is reported in Figure \ref{f9} as a full red line for the case $N=16$. As we can see, the agreement is satisfactory. Finally, from the determined $\alpha(\epsilon)$, using the condition $\Sigma_2(\epsilon_c)=\alpha(\epsilon_c)$, we found $\epsilon_c =0.835$ ($ \eta_c = 0.321 $), a value that is slightly higher (lower) than the one found with the assumption $\alpha(\epsilon)=\alpha(\epsilon=1)$.

\newpage

\section*{References}
\bibliographystyle{unsrt}
\bibliography{ref}

\begin{thebibliography}{10}

\bibitem{Sporns2013}
Olaf Sporns.
\newblock Network attributes for segregation and integration in the human
  brain.
\newblock {\em Current opinion in neurobiology}, 2013.

\bibitem{Park2013}
Hae-Jeong Park and Karl Friston.
\newblock Structural and functional brain networks: From connections to
  cognition.
\newblock {\em Science}, 342(6158):1238411, 2013.

\bibitem{Martin2001}
A.~Martin and L.~L. Chao.
\newblock Semantic memory and the brain: structure and processes.
\newblock {\em Cognitive Neuroscience}, 11:194--201, 2001.

\bibitem{Pascual2004}
Alberto Pascual, Kai-Lian Huang, Julie Neveu, and Thomas Pr\'eat.
\newblock Neuroanatomy: brain asymmetry and long-term memory.
\newblock {\em Nature}, 427(6975):605--6, Feb 2004.

\bibitem{Yamashita2015}
Masahiro Yamashita, Mitsuo Kawato, and Hiroshi Imamizu.
\newblock Predicting learning plateau of working memory from whole-brain
  intrinsic network connectivity patterns.
\newblock {\em Scientific Reports}, 5:7622, January 2015.

\bibitem{Shultz2016}
David Shultz.
\newblock Consciousness may be the product of carefully balanced chaos.
\newblock {\em Brain \& Behavior}, 2016.

\bibitem{Hayek1963}
Friedrich A.~Von Hayek.
\newblock {\em The Sensory Order: An Inquiry Into the Foundations of
  Theoretical Psychology}.
\newblock Martino Fine Books, 1963.

\bibitem{McCulloch1943}
Warren~S McCulloch and Walter Pitts.
\newblock A logical calculus of the ideas immanent in nervous activity.
\newblock {\em The Bulletin of Mathematical Biophysics}, 5(4):115--133, 1943.

\bibitem{Hopfield1982}
J.J. Hopfield.
\newblock Neural networks and physical systems with emergent collective
  computational abilities.
\newblock {\em Proceedings of the National Academy of Sciences},
  79(8):2554--2558, 1982.

\bibitem{Amit1985}
Daniel~J Amit, Hanoch Gutfreund, and Haim Sompolinsky.
\newblock Spin-glass models of neural networks.
\newblock {\em Physical Review A}, 32(2):1007, 1985.

\bibitem{Pfeiffer2015}
Brad~E. Pfeiffer and David~J. Foster.
\newblock Autoassociative dynamics in the generation of sequences of
  hippocampal place cells.
\newblock {\em Science}, 349(6244):180--183, 2015.

\bibitem{Fuster1971}
Joaquin~M Fuster, Garrett~E Alexander, et~al.
\newblock Neuron activity related to short-term memory.
\newblock {\em Science}, 173(3997):652--654, 1971.

\bibitem{Miyashita1988}
Y~Miyashita.
\newblock Neuronal correlate of visual associative long-term memory in the
  primate temporal cortex.
\newblock {\em Nature}, 335(6193):817--20, Oct 1988.

\bibitem{Tanaka1980}
F~Tanaka and S~F Edwards.
\newblock Analytic theory of the ground state properties of a spin glass. i.
  ising spin glass.
\newblock {\em Journal of Physics F: Metal Physics}, 10(12):2769, 1980.

\bibitem{Gutfreund1988}
H~Gutfreund, J~D Reger, and A~P Young.
\newblock The nature of attractors in an asymmetric spin glass with
  deterministic dynamics.
\newblock {\em Journal of Physics A: Mathematical and General}, 21(12):2775,
  1988.

\bibitem{Bastolla1998}
Ugo Bastolla and Giorgio Parisi.
\newblock Relaxation, closing probabilities and transition from oscillatory to
  chaotic attractors in asymmetric neural networks.
\newblock {\em Journal of Physics A: Mathematical and General}, 31(20):4583,
  1998.

\bibitem{Nutzel1991}
K~Nutzel.
\newblock The length of attractors in asymmetric random neural networks with
  deterministic dynamics.
\newblock {\em Journal of Physics A: Mathematical and General}, 24(3):L151,
  1991.

\bibitem{Molgedey1992}
L.~Molgedey, J.~Schuchhardt, and H.~G. Schuster.
\newblock Suppressing chaos in neural networks by noise.
\newblock {\em Phys. Rev. Lett.}, 69:3717--3719, Dec 1992.

\bibitem{Schuecker2016}
J.~{Schuecker}, S.~{Goedeke}, and M.~{Helias}.
\newblock {Optimal sequence memory in driven random networks}.
\newblock {\em ArXiv e-prints}, March 2016.

\bibitem{Tirozzi1991}
B.~Tirozzi and M.~Tsodyks.
\newblock Chaos in highly diluted neural networks.
\newblock {\em EPL (Europhysics Letters)}, 14(8):727, 1991.

\bibitem{Sompolinsky1988}
H.~Sompolinsky, A.~Crisanti, and H.~J. Sommers.
\newblock Chaos in random neural networks.
\newblock {\em Phys. Rev. Lett.}, 61:259--262, Jul 1988.

\bibitem{Crisanti1990}
H.~Sompolinsky A.~Crisanti, H.J.~Sommers.
\newblock Chaos in neural networks: chaotic solutions.
\newblock {\em preprint}, March 1990.

\bibitem{Stern2014}
M.~Stern, H.~Sompolinsky, and L.~F. Abbott.
\newblock Dynamics of random neural networks with bistable units.
\newblock {\em Phys. Rev. E}, 90:062710, Dec 2014.

\bibitem{Folli2017}
V.~Folli, G.~Gosti, M.~Leonetti, and G.~Ruocco.
\newblock Effect of dilution in asymmetric recurrent neural networks.
\newblock {\em Neural Networks}, 104:50, 2018.

\bibitem{GardnerE.1987}
{Gardner, E.}, {Derrida, B.}, and {Mottishaw, P.}
\newblock Zero temperature parallel dynamics for infinite range spin glasses
  and neural networks.
\newblock {\em J. Phys. France}, 48(5):741--755, 1987.

\bibitem{Derrida1986}
B.~Derrida and Y.~Pomeau.
\newblock Random networks of automata: A simple annealed approximation.
\newblock {\em EPL (Europhysics Letters)}, 1(2):45, 1986.

\bibitem{DerridaB.1987}
{Derrida, B.} and {Flyvbjerg, H.}
\newblock The random map model: a disordered model with deterministic dynamics.
\newblock {\em J. Phys. France}, 48(6):971--978, 1987.

\bibitem{Diestel2005}
Reinhard Diestel.
\newblock {\em Graph Theory (Graduate Texts in Mathematics)}.
\newblock Springer, 2005.

\bibitem{Bastolla1997}
U~Bastolla and G~Parisi.
\newblock {Attractors in fully asymmetric neural networks}.
\newblock {\em Journal of Physics A: Mathematical and General}, 30(16):5613,
  1997.

\end{thebibliography}

\end{document}